\documentclass[12pt]{article}
\usepackage{amssymb,amsmath,epsfig}
\allowdisplaybreaks
\begin{document}
\title{\bf Isotropic Exact Solutions via Noether Symmetries in $f(R,T,R_{\mu\nu}T^{\mu\nu})$ Gravity}

\author{Iqra Nawazish \thanks{iqranawazish07@gmail.com} and M. Sharif \thanks{msharif.math@pu.edu.pk}\\
Department of Mathematics and Statistics, The University of Lahore,\\
1.5 KM Defence Road Lahore-54000, Pakistan.}

\date{}

\maketitle
\begin{abstract}
In this paper, we study cosmic evolutionary stages in the background
of modified theory admitting non-minimal coupling between Ricci
scalar, trace of the energy-momentum tensor, contracted Ricci and
energy-momentum tensors. For dust distribution, we consider
isotropic, homogeneous and flat cosmic model to determine symmetry
generators, conserved integrals and exact solutions using Noether
symmetry scheme. We find maximum symmetries for non-minimally
interacting Ricci scalar and trace of the energy-momentum tensor but
none of them correspond to any standard symmetry. For rest of the
models, we obtain scaling symmetry with conserved linear momentum.
The graphical analysis of standard cosmological parameters, squared
speed of sound, viability conditions suggested by Dolgov-Kawasaki
instability and state-finder parameters identify realistic nature of
new models compatible with Chaplygin gas model, quintessence and
phantom regions. The fractional densities relative to ordinary
matter and dark energy are found to be consistent with Planck 2018
observational data. It is concluded that the constructed
non-minimally coupled models successfully explore cosmic accelerated
expansion.
\end{abstract}

\section{Introduction}

The theory of general relativity (GR) is based on the most
intellectual idea that connects curvature and matter contents to
interpret the geometry of spacetime and dynamics of gravitating
objects. Besides the attractive nature of gravity, modern cosmology
unravels the presence of an obscured repulsive force being
responsible for current cosmic expansion. The enigmatic description
of this force triggers cosmologists to put forward the most
compatible explanation of an exotic form of fluid with negative
pressure. This negative pressure is assumed to be the source of
strong anti-gravitational force whose striking features are still
unknown and consequently, referred to as dark energy (DE). To
interpret strong gravitational interactions and intriguing
characteristics of DE, GR requires few modifications that increase
differential order of the Einstein field equations and ensure the
existence of an extra force deviating massive particles from
geodesic to non-geodesic lines of motion \cite{1a}. The most
convincing approach is to redefine the geometric part of
Einstein-Hilbert action that supports minimal as well as non-minimal
coupling between geometric and matter contents. This approach
achieves a milestone on the road of theoretical advancements by
introducing modified gravitational theories like $f(R)$, $f(R,T)$
and $f(R,T,R_{\mu\nu}T^{\mu\nu})$ ($R_{\mu\nu}$, $T^{\mu\nu}$, $R$
and $T$, denote Ricci and energy-momentum tensors with their traces,
respectively) theories. The minimally coupled gravitational theories
follow equivalence principle, geodesic motion and preserve
conservation of energy-momentum tensor whereas the geometric
infrastructure of these theories unravel various cosmological
phenomenologies \cite{2}.

The non-minimal interactions of curvature and matter contents
significantly investigate early expansion, cosmological evolution
and current cosmic state in the background of both geodesic as well
as non-geodesic motion. Bertolami et al. \cite{3} followed this
concept of non-minimal interactions in $f(R)$ theory and found an
extra force that deviates massive test particles from their
geodesics. Harko et al. \cite{4} considered non-minimal coupling
between $R$ and $T$, and consequently proposed $f(R,T)$ theory.
Besides cosmic expansion and evolution, such advancements
successively explore realistic/unrealistic cosmological
configurations and possible explanations for early expansion
\cite{5}. In the framework of $f(R,T)$ theory, Sharif and Zubair
\cite{6} dealt with different cosmic issues of isotropic as well as
anisotropic cosmological models. In the last decade, researchers
have taken a keen interest to study thermodynamical laws, energy
constraints for different cosmological models, gravitational
collapse, dynamical instabilities, cosmic evolution and cosmological
solutions for self-gravitating objects in the framework of
$f(R,T,R_{\mu\nu}T^{\mu\nu})$ gravity \cite{ab}.

The evaluation of cosmological exact solutions provides a
significant way to understand evolution, configurations and matter
contribution in the universe. Sebastiani and Zerbini \cite{a1}
determined static spherically symmetric solutions for constant as
well as variable Ricci scalar in $f(R)$ theory. For minimally
interacting scalar field with Ricci scalar, Maharaj et al.
\cite{a12} restricted constant of integration to find de Sitter,
oscillating, accelerating, decelerating and contracting cosmological
solutions. Sharif and Zubair \cite{11} measured anisotropic
solutions corresponding to power-law and exponential cosmological
models in $f(R,T)$ theory. Harko and Lake \cite{8} used particular
constraints to calculate cylindrical solutions in the same theory.
Shamir \cite{12} considered anisotropic Bianchi I cosmological model
and obtained three unique solutions whose physical behavior is
analyzed via standard cosmological parameters. Besides evaluating
exact solutions, different authors \cite{ref1} studied dynamical
features of cosmological configurations in Gauss-Bonnet gravity and
Einstein-Maxwell scalar theory.

Capozziello et al. \cite{20} used Noether symmetry technique to
evaluate static spherically symmetric solution with constant Ricci
scalar and power-law $f(R)$ model. Hussain et al. \cite{23} measured
symmetries for flat, isotropic and homogeneous cosmological model
with the same $f(R)$ model. Shamir et al. \cite{24} used Noether
gauge symmetry condition for power-law $f(R)$ isotropic cosmological
model and obtained some extra symmetries. $\ddot{O}$z and Bamba
\cite{24a} discovered some new solutions for static cylindrical and
planar spacetimes with non-dust fluid and same $f(R)$ model.
Atazadeh and Darabi \cite{15} solved equations of motion via Noether
symmetry technique and obtained some viable $f(\mathcal{T})$ models
($\mathcal{T}$ denotes torsion) that favor power-law expansion.
Momeni et al. \cite{a3} constructed an over-determining system of
non-linear equations in the framework of mimetic $f(R)$ gravity that
yields Noether point symmetries whereas they also determined
power-law solution supporting decelerating cosmic expansion in
$f(R,T)$ theory. Sharif and his collaborators \cite{a2} studied
cosmic evolution by establishing symmetry generators and conserved
entities of isotropic as well as anisotropic cosmological models in
the background of $f(R)$, $f(R,T)$ and $f(\mathcal{G})$ theories
($\mathcal{G}$, referred to Gauss-Bonnet invariant). Besides
exploring current cosmic expansion and evolution, they also
determined the existence of cosmological configurations like
wormholes whose realistic nature and structure are examined via
stability constraints, geometric conditions and energy bounds in the
same theories \cite{16}. In $f(R,\phi,X)$ ($\phi$ and $X$ specify
scalar field and kinetic term of scalar field, respectively) theory,
different researchers solved isotropic, anisotropic and static
spherically symmetric cosmological models via Noether symmetry
scheme \cite{16a}.

The evaluation of new cosmological models and exact solutions
significantly lead the way to understand interaction of matter with
geometry and its contribution in the universe. In non-minimally
coupled theories, it is a bit of task to determine exact solutions
of non-linear higher order field equations. This problem compelled
to think of some other ways that will reduce the complexity of these
equations as well as lead to construct exact solutions. In order to
understand the impact of strong non-minimal curvature-matter
interactions on current cosmos, we consider flat isotropic dust
cosmological model and use Noether symmetry approach to calculate
new cosmological models. We determine symmetries as well as relative
conserved integrals in $f(R,T,R_{\mu\nu}T^{\mu\nu})$ theory.

The layout of the paper is given as follows. The basic background of
this theory and key steps of this approach are provided in section
\textbf{2}. In sections \textbf{3-6}, we obtain Noether symmetry,
conserved quantities and corresponding exact solutions. We establish
graphical analysis of standard cosmological parameters and
fractional densities to understand the behavior of solutions whereas
model consistency is investigated via Dolgov-Kawasaki instability
constraints, squared speed of sound and state-finder parameters. In
the last section, we provide a detailed summary of our results.

\section{Background of $f(R,T,R_{\mu\nu}T^{\mu\nu})$ Gravity
and Noether Symmetry Approach}

Odintsov and Saez-Gomez \cite{b1} considered non-minimally coupled
contracted Ricci and energy-momentum tensors in $f(R,T)$ theory and
extended this theory to $f(R,T,R_{\mu\nu}T^{\mu\nu})$ gravitational
theory. The Einstein-Hilbert action admitting non-minimal
interaction among Ricci scalar, trace of the energy-momentum tensor
with contracted Ricci and energy-momentum tensors is defined as
\begin{equation}\label{1}
\mathcal{A}=\frac{1}{2\kappa^2}\int
\sqrt{-g}f(R,T,R_{\mu\nu}T^{\mu\nu})d^4x+
\int\sqrt{-g}\mathcal{L}_md^4x.
\end{equation}
Here the first integral defines geometric Lagrangian depending on a
generic function $f$ that incorporates non-minimal coupling between
curvature and matter variables whereas $\kappa$ represents coupling
constant and $g$ describes determinant of the metric tensor
($g_{\mu\nu}$). The second integral related to ordinary matter
Lagrangian density is denoted as $\mathcal{L}_m$. The matter
Lagrangian density is independent of first order derivative of the
metric tensor and consequently, leads to the following
energy-momentum tensor
\begin{equation}\label{2}
T_{\mu\nu}=-\frac{2}{\sqrt{-g}}\frac{\delta(\sqrt{-g}\mathcal{L}_m)}{\delta
g^{\mu\nu}}=g_{\mu\nu}\mathcal{L}_{m}
-\frac{2\delta\mathcal{L}_{m}}{\delta g^{\mu\nu}}.
\end{equation}
For $\kappa^2=1$ and $R_{\mu\nu}T^{\mu\nu}=Q$, the metric variation
of the action (\ref{1}) yields
\begin{eqnarray}\nonumber
&&(f_{_R}-f_{_Q}\mathcal{L}_{m})G_{\mu\nu} + \left[\Box
f_{_R}+\frac{1}{2}Rf_{_R}-\frac{1}{2}f+f_{_T}\mathcal{L}_{m}+
\frac{1}{2}\nabla_{\alpha}\nabla_{\beta}\left(f_{_Q}T^{\alpha\beta}
\right)\right]g_{\mu\nu}\\\nonumber
&&-\nabla_{\mu}\nabla_{\nu}f_{_R}
+\frac{1}{2}\Box(f_{_Q}T_{\mu\nu})+2f_{_Q}R_{\alpha(\mu}T_{\nu)}^{\alpha}
-\nabla_{\alpha}\nabla_{(\mu}[T_{\nu)}^{\alpha}f_{_Q}] -2
(f_{_T}g^{\alpha\beta}\\\label{3}&&+f_{_Q}R^{\alpha\beta}
)\frac{\partial^{2}\mathcal{L}_{m}}{\partial g^{\mu\nu}\partial
g^{\alpha\beta}} =(1+f_{_T}+\frac{1}{2}Rf_{_Q})T_{\mu\nu},
\end{eqnarray}
where $f_{_i}$, $i=R,~T,~Q$ represents derivative of the generic
function with respect to corresponding variable while $G_{\mu\nu}$
and $\nabla_{\mu}$ denote Einstein tensor and covariant derivative,
respectively and $\Box=\nabla_{\mu}\nabla^{\mu}$.

An equivalent expression for the above field equations is given by
\begin{equation}\label{4}
R_{\mu\nu}-\frac{1}{2}Rg_{\mu\nu}=T_{\mu\nu}^{eff}=T_{\mu\nu}^{c}+T_{\mu\nu},
\end{equation}
where $T_{\mu\nu}^{c}$ defines energy-momentum tensor relative to
higher-order non-linear curvature terms whereas the effective
energy-momentum tensor $T_{\mu\nu}^{eff}$ incorporates a combination
of ordinary matter variables and curvature terms given as
\begin{eqnarray}\nonumber
T_{\mu\nu}^{eff}&=&\frac{1}{f_{_R}-f_{_Q}\mathcal{L}_{m}}[(1+f_{_T}+\frac{1}{2}
Rf_{_Q})T_{\mu\nu}+\{\frac{1}{2}(f-Rf_{_R})-\mathcal{L}_{m}f_{_T}\\\nonumber&-&
\frac{1}{2}\nabla_{\alpha}\nabla_{\beta}(f_{_Q}T^{\alpha\beta})\}g_{\mu\nu}
-(g_{\mu\nu}\Box-\nabla_{\mu}\nabla_{\nu})f_{_R}-\frac{1}{2}\Box(f_{_Q}T_{\mu\nu})
\\\label{5}&+&\nabla_{\alpha}\nabla_{(\mu}[T_{\nu)}^{\alpha}f_{_Q}]-2f_{_Q}
R_{\alpha(\mu}T_{\nu)}^{\alpha}+2(f_{_T}g^{\alpha\beta}+f_{_Q}R^{\alpha\beta})
\frac{\partial^{2}\mathcal{L}_{m}}{\partial g^{\mu\nu}\partial
g^{\alpha\beta}}].
\end{eqnarray}
From the metric contraction of Eq.(\ref{3}), we obtain a significant
equation that preserves relationship between traces of geometric and
matter parts as follows
\begin{eqnarray*}\nonumber
&&\nabla_{\alpha}\nabla_{\beta}(f_{_Q}T^{\alpha\beta})+(f_{_R}+f_{_Q}\mathcal{L}_m)R
+4f_{_T}\mathcal{L}_m-2f+3\Box
f_{_R}+\frac{1}{2}\Box(f_{_Q}T)\\\nonumber&&+2f_{_Q}R_{\alpha\beta}T^{\alpha\beta}-2g^{\mu\nu}
(f_{_T}g^{\alpha\beta}+f_{_Q}R^{\alpha\beta})\frac{\partial^{2}\mathcal{L}_{m}}{\partial
g^{\mu\nu}\partial g^{\alpha\beta}}=(1+f_{_T}+\frac{1}{2}Rf_{_Q})T.
\end{eqnarray*}
In the present case, the non-conserved effective energy-momentum
tensor takes the following form
\begin{eqnarray}\nonumber
&&\nabla^{\mu}T_{\mu\nu}=\frac{1}{1+f_{_T}+(Rf_{_Q})/2}\left[\nabla_{\mu}
(f_{_Q}T_{\sigma\nu}R^{\sigma\mu})-\frac{1}{2}(f_{_Q}f_{_T}R_{\alpha\beta}
g_{\alpha\beta})\nabla_{\nu}T^{\alpha\beta}\right.\\\label{T}&&\left.
+\nabla_{\nu}(\mathcal{L}_mf_{_T})
-G_{\mu\nu}\nabla^{\mu}(f_{_Q}\mathcal{L}_m)-\frac{1}{2}(\nabla^{\mu}(Rf_{_Q})
+2\nabla^{\mu}f_{_T})T_{\mu\nu}\right].
\end{eqnarray}
This non-conserved energy-momentum tensor appears due to the
existence of an additional force (${\mathcal{F}}^{\alpha}$)
perpendicular to four velocity of the massive particles given by
\begin{eqnarray*}\label{T}
&&\frac{d^2x^{\alpha}}{ds^2}+\Gamma^{\alpha}_{\mu\nu}u^{\mu}u^{\nu}=
\mathcal{F}^{\alpha}.
\end{eqnarray*}
In $f(R,T,Q)$ theory, this extra force takes the following form
\begin{eqnarray}\nonumber
&&\mathcal{F}^{\alpha}=\frac{h^{\alpha\beta}}
{(\rho_m+p_m)(1+2f_T+Rf_{_{RT}})}[(f_T+Rf_{_{RT}})
\nabla_{\beta}\rho_m-(1+3f_T)\nabla_{\beta}p_m\\\label{T1}&&-(\rho_m+p_m)f_{_{RT}}
R^{\gamma\delta}(\nabla\beta h_{\gamma\delta}-2\nabla\delta
h_{\gamma\beta})-f_{_{RT}}R_{\gamma\delta}h^{\gamma\delta}
\nabla_{\beta}(\rho_m+p_m)],
\end{eqnarray}
where $h^{\alpha\beta}$ represents projection vector.

The compatibility between minimally and non-minimally coupled
theories could be established if the equivalence principle and
conservation of effective energy-momentum tensor are preserved. In
case of $f(R,T)$ theory, this consistency can be achieved for
perfect fluid distribution with $f_{T}(R,T)=0$ as additional force
becomes perpendicular to four velocity. For pressureless fluid, the
effect of additional force can be avoided even with
$f_{T}(R,T)\neq0$. For $f(R,T,Q)$ gravitational theory, this extra
force cannot be neglected even for pressureless fluid due to its
dependence on the Ricci tensor. This dependence significantly
interprets the impact of interacting matter and curvature components
on cosmological configurations, their structures, thermodynamical
features and cosmic evolution. In the present case, the non-geodesic
motion of test particles can be reduced to geodesic motion only for
non-interacting curvature invariant and matter variables, i.e.,
$f_{T}(R,T,Q)=f_{_{RT}}=0$ \cite{4}.

The contribution of matter Lagrangian acts as stepping stone to
explore conserved or non-conserved nature of matter. In order to
understand the behavior of geodesic as well as non-geodesic motion
of massive test particles, one can freely choose the Lagrangian
density to be pressure or energy density dependent due to its
non-unique behavior. For pressure dependant matter Lagrangian
density ($\mathcal{L}_m=p_m$, $p_m$ denotes pressure of normal
matter), the extra force vanishes in $f(R,\mathcal{L}_m)$ theory
\cite{b2}. The universe is considered to be distributed with perfect
fluid whose energy-momentum is defined as follows
\begin{equation*}\label{F}
T_{\mu\nu}=(\rho_m+p_m)u_\mu u_\nu+p_mg_{\mu\nu},\quad
u_\mu=(-1,0,0,0),
\end{equation*}
where $\rho_m$ refers to energy density and $u_\mu$ specifies four
velocity of co-moving frame. The choice of matter Lagrangian does
not affect the non-conserved nature of effective energy-momentum
tensor as the additional force deviates test particles from
geodesics even for $\mathcal{L}_m=-\rho_m$ or $\mathcal{L}_m=p_m$.
Therefore, we can freely choose $\mathcal{L}_m=-\rho_m$ for perfect
fluid distribution.

The cosmological model for flat isotropic homogenous universe is
given as
\begin{equation}\label{6a}
ds^2=-dt^2+a^2(t)(dx^2+dy^2+dz^2),
\end{equation}
where $a(t)$ represents the scale factor measuring cosmic expansion
in spatial directions. In non-minimally coupled theories, the
Lagrange multiplier approach is helpful to construct point-like
Lagrangian. Using this approach, the action (\ref{1}) leads to the
following form
\begin{eqnarray}\nonumber
\mathcal{I}&=&\int\sqrt{-g}[f(R,T,Q)-\eta(R-\bar{R})-\zeta(T-\bar{T})
-\varphi(Q-\bar{Q})+\rho_m(a)]dt,\\\label{6}\bar{R}&=&6\left(\frac{\ddot{a}}{a}
+\frac{\dot{a^2}}{a^2}\right),\quad\bar{T}=3p_m-\rho_m,\quad
\bar{Q}=-\frac{3\ddot{a}\rho_m}{a}+\frac{3p_m}{a^2}\left(2\dot{a}^2+a\ddot{a}\right).
\end{eqnarray}
Here $\sqrt{-g}=a^3$, $\eta,~\zeta$ and $\varphi$ are scalar terms
while $\bar{R},~\bar{T}$ and $\bar{Q}$ define dynamical constraints.
Varying the above action with respect to $R,~T$ and $Q$, we get
$\eta=f_{_R},~\zeta=f_{_T}$ and $\varphi=f_{_Q}$, respectively. In
order to eliminate second-order derivatives in Eq.(\ref{6}), we
integrate the action by parts that yield
\begin{eqnarray}\nonumber
&&\mathcal{L}(a,R,T,Q,\dot{a},\dot{R},\dot{T},\dot{Q})=a^3[f(R,T,Q)-Rf_{_R}
-Tf_{_T}-Qf_{_Q}-f_{_T}(\rho_m-3p_m)\\\nonumber&&-\rho_m]-6a\dot{a}^2f_{_R}
-6a^2\dot{a}\dot{R}f_{_{RR}}-6a^2\dot{a}\dot{T}f_{_{RT}}
-6a^2\dot{a}\dot{Q}f_{_{RQ}}+3a\dot{a}^2f_{_Q}(2\rho_m-ap_m'\\\label{7}&&
+a\rho_m')+3a^2\dot{a}(\rho_m-p_m)[\dot{R}f_{_{RQ}}+\dot{T}f_{_{TQ}}
+\dot{Q}f_{_{QQ}}].
\end{eqnarray}

The Hamiltonian and Euler-Lagrange equations are significantly
helpful to determine total energy and equations of motion of a
dynamical system. The mathematical form of these equations is given
as
\begin{eqnarray}\nonumber
\mathcal{H}=\Sigma_i\dot{q}^i\mathcal{P}_i-\mathcal{L},\quad
\frac{\partial\mathcal{L}}{\partial
q^i}-\frac{d\mathcal{P}_i}{dt}=0,\quad\mathcal{P}_i=\frac{\partial\mathcal{L}}{\partial
\dot{q}^i},
\end{eqnarray}
where $q^i,~\mathcal{P}^i$ and $\dot{q}^i$ represent generalized
coordinates, momentum and velocity of the dynamical system,
respectively. For Eq.(\ref{7}), the Hamiltonian equation turns out
to be
\begin{eqnarray}\nonumber
\mathcal{H}&=&-\frac{3\dot{a^2}}{a^2}+\frac{1}{2(f_{_R}-\rho_mf_{_Q})}\left[f(R,T,Q)
-Rf_{_R}-Tf_{_T}-f_T(\rho_m-3p_m)\right.\\\nonumber&+&\rho_m-Qf_{_Q}+\left.3\dot{a}H
(\rho_m'-p_m')f_{_Q}+3H(\rho_m-p_m)\{\dot{R}f_{_{RQ}}
+\dot{T}f_{_{TQ}}+\dot{Q}f_{_{QQ}}\}\right.\\\label{8}&-&\left.6H\{\dot{R}f_{_{RR}}
+\dot{T}f_{_{RT}}+\dot{Q}f_{_{RQ}}\}\right].
\end{eqnarray}
The Hamiltonian equation also evaluates total energy density for
constraint $\mathcal{H}=0$. For generalized co-ordinates
$q_i=\{a,R,T,Q\}$, the corresponding Euler-Lagrangian equations
become
\begin{eqnarray}\nonumber
&&\dot{a^2}+2a\ddot{a}+\frac{a^2}{2(f_{_R}-\rho_mf_{_Q})}\left[f(R,T,Q)
-Rf_{_R}-Tf_{_T}-Qf_{_Q}-f_T(\rho_m-3p_m)\right.\\\nonumber&&-\left.\rho_m-
\frac{a}{3}(f_T(\rho_m'-3p_m')+\rho_m')+2H\{2\dot{f_{_R}}
-\dot{f_{_Q}}(2\rho_m+a\rho_m'-ap_m')\}\right.\\\nonumber&&+\left.2(\ddot{R}f_{_{RR}}+
\ddot{T}f_{_{RT}}+\ddot{Q}f_{_{RQ}}+\dot{R}\dot{f_{_{RR}}}+
+\dot{T}\dot{f_{_{RT}}}+\dot{Q}\dot{f_{_{RQ}}})+(\ddot{a}+2aH^2)
\right.\\\nonumber&&\times\left.
(\rho_m'-p_m')f_{_Q}-aH^2f_{_Q}(6\rho_m'-2p_m'+a\rho_m''-ap_m'')
+(\rho_m-p_m)\right.\\\label{9}&&\times\left.(\ddot{R}f_{_{RQ}}+
\ddot{T}f_{_{TQ}}+\ddot{Q}f_{_{QQ}}+\dot{R}\dot{f_{_{RQ}}}+
+\dot{T}\dot{f_{_{TQ}}}+\dot{Q}\dot{f_{_{QQ}}})\right]=0,
\\\nonumber&&-a^3(Rf_{_{RR}}+Tf_{_{RT}}+Qf_{_{RQ}}+f_{_{RT}}(\rho_m-3p_m))
+6a\dot{a}^2(f_{_{RR}}+\rho_mf_{_{RQ}})\\\label{10}&&
+6a^2\ddot{a}f_{_{RR}}-3af_{_{RQ}}(\rho_m-p_m)(2\dot{a}^2+a\ddot{a})=0,
\\\nonumber&&-a^3(Rf_{_{RT}}+Tf_{_{TT}}+Qf_{_{TQ}}+f_{_{TT}}(\rho_m-3p_m))
+6a\dot{a}^2(f_{_{RT}}+\rho_mf_{_{TQ}})\\\label{11}&&
+6a^2\ddot{a}f_{_{RT}}-3af_{_{TQ}}(\rho_m-p_m)(2\dot{a}^2+a\ddot{a})=0,
\\\nonumber&&-a^3(Rf_{_{RQ}}+Tf_{_{TQ}}+Qf_{_{QQ}}+f_{_{TQ}}(\rho_m-3p_m))
+6a\dot{a}^2(f_{_{RQ}}+\rho_mf_{_{QQ}})\\\label{12}&&
+6a^2\ddot{a}f_{_{RQ}}-3af_{_{QQ}}(\rho_m-p_m)(2\dot{a}^2+a\ddot{a})=0.
\end{eqnarray}

The exact solutions of the above non-linear partial differential
equations can be determined via Noether symmetry approach that
remarkably minimizes the complexity of dynamical equations
\cite{b4}. The formalism depends on a well-known Noether principle
which states that the invariance of Lagrangian along a vector field
establishes a relationship between symmetries induced by symmetry
generators and conservation. A gravitational theory admitting
minimal or non-minimal coupling is referred to as physically viable
if it retains symmetries and conserved entities. For affine
parameter $t$ and generalized co-ordinates $q^i$, the vector field
with invariance condition and Noether first integral are defined
mathematically as follows
\begin{eqnarray}\nonumber
Y&=&\chi(t,q^i)\partial_t+\xi^j(t,q^i)\partial_{q^j},\quad
Y^{[1]}\mathcal{L}+(D\chi)\mathcal{L}=D\mathcal{B}(t,q^i),\\\label{13}
I&=&\mathcal{B}-\chi\mathcal{L}-(\xi^j-\dot{q}^j\chi)
\frac{\partial\mathcal{L}}{\partial\dot{q}^j},
\end{eqnarray}
where $\chi$ and $\xi^j$ are unknown functions depending on affine
parameter and canonical variables $q^j=a,R,T,Q$. The boundary term
($\mathcal{B}$) identifies some extra symmetries that are referred
to as Noether gauge symmetry while the first order prolongation
$Y^{[1]}$ and total derivative $D$ are given by
\begin{eqnarray}\label{15}
Y^{[1]}=Y+(\chi^j,_t+\chi^j,_i\dot{q}^i-\xi,_t\dot{q}^j
-\xi,_i\dot{q}^i\dot{q}^j)\frac{\partial}{\partial\dot{q}^j},\quad
D=\partial_t+\dot{q}^i\partial_{q^i}.
\end{eqnarray}

Due to the absence of boundary term ($\mathcal{B}=0$), the first
order prolongation becomes zero as the unknown coefficients are
independent of affine parameter. For this scenario, the Noether
gauge symmetry reduces into simple Noether symmetry and
consequently, Eq.(\ref{13}) becomes
\begin{equation}\label{17}
Y=\xi^i(q^i)\partial_{q^i}+\left[\frac{d}{dt}(\xi^i(q^i))\right]\partial_{\dot{q}^i},
\quad L_Y\mathcal{L}=Y\mathcal{L}=0,\quad
I=-\xi^j\frac{\partial\mathcal{L}}{\partial\dot{q}^j},
\end{equation}
where $L$ specifies Lagrangian's Lie derivative. For the
configuration $\mathcal{Q}=\{a,R,T,Q\}$ and tangent space
$\mathcal{T}=\{a,\dot{a},R,\dot{R},T,\dot{T},Q,\dot{Q}\}$, the
corresponding vector field takes the following form
\begin{equation}\label{46}
Y=\alpha\partial_{_{a}}+\beta\partial_{_{R}}+\phi
\partial_{_{T}}+\psi\partial_{_{Q}}+\dot{\alpha}\partial_{_{\dot{a}}}
+\dot{\beta}\partial_{_{\dot{R}}}+\dot{\phi}\partial_{_{\dot{T}}}
+\dot{\psi}\partial_{_{\dot{Q}}}.
\end{equation}
The time derivatives of the above unknown coefficients are given by
\begin{eqnarray}\nonumber
\dot{\alpha}&=&\alpha_{_t}+\dot{a}\alpha_{_a}
+\dot{R}\alpha_{_R}+\dot{T}\alpha_{_T}
+\dot{Q}\alpha_{_Q},\quad\dot{\beta}=\beta_{_t}+\dot{a}\beta_{_a}
+\dot{R}\beta_{_R}+\dot{T}\beta_{_T}
+\beta\alpha_{_Q},\\
\nonumber \dot{\phi}&=&\phi_{_t}+\dot{a}\phi_{_a}
+\dot{R}\phi_{_R}+\dot{T}\phi_{_T}
+\dot{Q}\phi_{_Q},\quad\dot{\psi}=\psi_{_t}+\dot{a}\psi_{_a}
+\dot{R}\psi_{_R}+\dot{T}\psi_{_T} +\dot{Q}\psi_{_Q}.
\end{eqnarray}
The invariance condition (\ref{17}) for the vector field (\ref{46})
yields
\begin{eqnarray}\label{20a}
&&\alpha_{_{R}}[2f_{_{RR}}-(\rho_m-p_m)f_{_{RQ}}]=0,
\\\label{21aa}&&\alpha_{_{T}}[2f_{_{RT}}-(\rho_m-p_m)f_{_{TQ}}]=0,
\\\label{21a}&&\alpha_{_{Q}}[2f_{_{RQ}}-(\rho_m-p_m)f_{_{QQ}}]=0,
\\\label{29a}&&\alpha_{_{R}}[2f_{_{RT}}-(\rho_m-p_m)f_{_{TQ}}]
+\alpha_{_{T}}[2f_{_{RR}}-(\rho_m-p_m)f_{_{RQ}}]=0,
\\\label{30a}&&\alpha_{_{R}}[2f_{_{RQ}}-(\rho_m-p_m)f_{_{QQ}}]
+\alpha_{_{Q}}[2f_{_{RR}}-(\rho_m-p_m)f_{_{RQ}}]=0,
\\\label{31a}&&\alpha_{_{Q}}[2f_{_{RT}}-(\rho_m-p_m)f_{_{TQ}}]
+\alpha_{_{T}}[2f_{_{RQ}}-(\rho_m-p_m)f_{_{QQ}}]=0,
\\\nonumber&&\alpha(a^2f_{_{Q}}(\rho_m''-p_m'')-2(f_{_{R}}
-af_{_{Q}}(\rho_m'-p_m')-f_{_{Q}}\rho_m))+a\beta(-2f_{_{RR}}
+f_{_{RQ}}\\\nonumber&&\times(2\rho_m+a\rho_m'-ap_m'))+a\phi(-2f_{_{RT}}
+(2\rho_m+a\rho_m'-ap_m')f_{_{TQ}})+a\psi(-2f_{_{RQ}}\\\nonumber&&
+(2\rho_m+a\rho_m'-ap_m')f_{_{QQ}})+2a\alpha_{_{a}}((2\rho_m+a\rho_m'-ap_m')
f_{_{Q}}-2f_{_{R}})-a^2\beta_{_{a}}\\\nonumber&&\times(2f_{_{RR}}
-(\rho_m-p_m')f_{_{RQ}})-a^2\phi_{_{a}}(2f_{_{RT}}
-(\rho_m-p_m')f_{_{TQ}})-a^2\psi_{_{a}}(2f_{_{RQ}}\\\label{40a}&&
-(\rho_m-p_m)f_{_{QQ}})=0,\\\nonumber&&
\alpha(-4f_{_{RR}}+af_{_{RQ}}(\rho_m'-p_m')+2(\rho_m-p_m)f_{_{RQ}})
-a\beta(2f_{_{RRR}}-f_{_{RRQ}}(\rho_m-p_m))\\\nonumber&&
-a\phi(2f_{_{RRT}}-(\rho_m-p_m)f_{_{RTQ}})
-a\psi(2f_{_{RRQ}}-(\rho_m-p_m)f_{_{RQQ}})-a\alpha_{_{a}}
(2f_{_{RR}}\\\nonumber&&-f_{_{RQ}}(\rho_m-p_m))+2\alpha_{_{R}}((2\rho_m+a\rho_m'-ap_m')
f_{_{Q}}-2f_{_{R}})-a\beta_{_{R}}(2f_{_{RR}}
-f_{_{RQ}}\\\nonumber&&\times(\rho_m-p_m))-a\phi_{_{R}}(2f_{_{RT}}
-(\rho_m-p_m)f_{_{TQ}})-a\psi_{_{R}}(2f_{_{RQ}}
-(\rho_m-p_m)f_{_{QQ}})\\\label{41a}&&=0,\\\nonumber&&
\alpha(-4f_{_{RT}}+af_{_{TQ}}(\rho_m'-p_m')+2(\rho_m-p_m)f_{_{TQ}})
-a\beta(2f_{_{RRT}}-f_{_{RTQ}}(\rho_m-p_m))\\\nonumber&&
-a\phi(2f_{_{RTT}}-(\rho_m-p_m)f_{_{TTQ}})
-a\psi(2f_{_{RTQ}}-(\rho_m-p_m)f_{_{TQQ}})-a\alpha_{_{a}}
(2f_{_{RT}}\\\nonumber&&-f_{_{TQ}}(\rho_m-p_m))+2\alpha_{_{T}}((2\rho_m+a\rho_m'-ap_m')
f_{_{Q}}-2f_{_{R}})-a\beta_{_{T}}(2f_{_{RR}}
-f_{_{RQ}}\\\nonumber&&\times(\rho_m-p_m))-a\phi_{_{T}}(2f_{_{RT}}
-(\rho_m-p_m)f_{_{TQ}})-a\psi_{_{T}}(2f_{_{RQ}}
-(\rho_m-p_m)f_{_{QQ}})\\\label{42a}&&=0,\\\nonumber&&
\alpha(-4f_{_{RQ}}+af_{_{QQ}}(\rho_m'-p_m')+2(\rho_m-p_m)f_{_{QQ}})
-a\beta(2f_{_{RRQ}}-f_{_{RQQ}}(\rho_m-p_m))\\\nonumber&&
-a\phi(2f_{_{RTQ}}-(\rho_m-p_m)f_{_{TQQ}})
-a\psi(2f_{_{RQQ}}-(\rho_m-p_m)f_{_{QQQ}})-a\alpha_{_{a}}
(2f_{_{RQ}}\\\nonumber&&-f_{_{QQ}}(\rho_m-p_m))+2\alpha_{_{Q}}((2\rho_m+a\rho_m'-ap_m')
f_{_{Q}}-2f_{_{R}})-a\beta_{_{Q}}(2f_{_{RR}}
-f_{_{RQ}}\\\nonumber&&\times(\rho_m-p_m))-a\phi_{_{Q}}(2f_{_{RT}}
-(\rho_m-p_m)f_{_{TQ}})-a\psi_{_{Q}}(2f_{_{RQ}}
-(\rho_m-p_m)f_{_{QQ}})\\\label{43a}&&=0,\\\nonumber&&
\alpha(3a^2\{f-Rf_{_{R}}-Tf_{_{T}}-Qf_{_{Q}}-f_{_{T}}(\rho_m-3p_m)-\rho_m\}
-a^3\{f_{_{T}}(\rho_m'\\\nonumber&&-3p_m')-\rho_m'\})+a^3\beta(-Rf_{_{RR}}-Tf_{_{RT}}
-Qf_{_{RQ}}-(\rho_m-3p_m)f_{_{RT}})+a^3\phi\\\nonumber&&\times(-Rf_{_{RT}}-Tf_{_{TT}}
-Qf_{_{TQ}}-(\rho_m-3p_m)f_{_{TT}})+a^3\psi(-Rf_{_{RQ}}-Tf_{_{TQ}}\\\label{39a}&&
-Qf_{_{QQ}}-(\rho_m-3p_m)f_{_{TQ}})=0.
\end{eqnarray}

In order to study the impact of geodesic as well as non-geodesic
strong curvature regimes in the cosmos, we consider different
possibilities for interactions among curvature scalar, trace of the
energy-momentum tensor, contracted Ricci and energy-momentum
tensors. For pressureless perfect fluid ($p_m=0$), the non-geodesic
equation of motion recovers standard geodesic equation with $T=0$,
$Q=0$, $f_T(R,T)=0$ and $f_{_{RT}}(R,T)=0$. These constraints lead
to define non-minimal models of $f(R,T,Q)$ theory such as $f(R,Q)$
and $f(R,T)$ models that appreciate direct interactions of Ricci
scalar with contracted Ricci, energy-momentum tensors and trace of
the energy-momentum tensor, respectively. For $f_T(R,T)\neq0$, we
consider $f(T,Q)$ model with $R=0$ to analyze cosmic evolution in
the presence of non-geodesic dust particles. Apart from these
non-minimally coupled models, it would be interesting to investigate
the crucial behavior of extra force in the presence of all three
variables, i.e., $R, T$ and $Q$. For the considered possibilities of
the generic function, we have
\begin{itemize}
\item $f(R,Q)$ model, independent of $T$,
\item $f(R,T)$ model, independent of $Q$,
\item $f(T,Q)$ model, independent of $R$,
\item $f(R,T,Q)$ model, with $R,~T,~Q\neq0$.
\end{itemize}
Our aim is to solve the above non-linear system of equations and
determine the existence of Noether point symmetries, associated
conserved entities that further help to evaluate exact solutions in
the background of dust fluid.

\section{Noether Symmetries of \textbf{$f(R,Q)$ Model}}

Here we consider generic function as the source of non-minimal
interactions between curvature scalar and contracted Ricci as well
as energy-momentum tensors while the contribution of trace of the
energy-momentum tensor is zero. For this choice of model, we explore
existence of Noether point symmetries that lead to conservation laws
and exact solutions. We also establish graphical analysis of some
standard cosmological parameters to understand the nature of
determined solutions. In this regard, we choose $\alpha=c_1$ to
solve the system of non-linear equations (\ref{20a})-(\ref{39a}) and
obtain
\begin{eqnarray}\nonumber
\alpha&=&c_1,\quad\beta=c_5c_6(c_7R+c_8)\frac{Z_1(a)Z_2(Q)}{Q},
\quad\phi=0,\quad\rho_m(a)=Z_1(a),
\\\label{47}\delta&=&c_5c_6(c_7R+c_8)\frac{Z_1(a)Z_2(Q)}{R},
\quad f(R,Q)=c_3c_4Q\left(\frac{R}{Q}\right)^{c_2}.
\end{eqnarray}
The system of equations satisfies these solutions for
$Z_1(a)=\frac{c_9}{a^2}$ and $Z_2(Q)=c_{10}Q$ whereas
$c_i~(i=1,2,...,10$) are constants. In this case, Noether point
symmetries and associated conserved entities are given as
\begin{eqnarray}\nonumber
Y_1&=&\partial_{_a},\quad
I_1=12a\dot{a}f_{_{R}}+6a^2(\dot{R}f_{_{RR}}+\dot{Q}f_{_{RQ}})
-3a^2\rho_m(\dot{R}f_{_{RQ}}+\dot{Q}f_{_{QQ}})
\\\nonumber&-&6a\dot{a}f_{_{Q}}(2\rho_m+a\rho_m'),\\\nonumber
Y_2&=&\frac{R}{a^2}\partial_{_{R}}+\frac{RQ}{a^2}\partial_{_{Q}},\quad
I_2=6\dot{a}R(f_{_{RR}}+Qf_{_{RQ}})-\frac{3c_9\dot{a}R}{a^2}
(f_{_{RQ}}+Qf_{_{QQ}}),\\\label{48}
Y_3&=&\frac{1}{a^2}(\partial_{_{R}}+Q\partial_{_{Q}}),\quad
I_3=6\dot{a}(f_{_{RR}}+f_{_{RQ}})-\frac{3c_9\dot{a}}{a^2}
(f_{_{RQ}}+f_{_{QQ}}).
\end{eqnarray}
The symmetry generator $Y_1$ and respective conserved integral $I_1$
refer to scaling symmetry and conserved linear momentum,
respectively. To construct cosmological analysis of $f(R,Q)$ model
(\ref{47}), we solve Eq.(\ref{48}) for $I_3=0$ and determine the
following exact solution
\begin{eqnarray}\nonumber
a(t)&=&\sqrt{(c_{11}t+c_{12})^2+c_9}.
\end{eqnarray}

To understand cosmological nature of this exact solution, we study
behavior of some important cosmological parameters graphically. We
consider Hubble ($H$), deceleration ($q$) and effective equation of
state ($w_{eff}$) parameters that measure rate of expansion and
further characterize it into accelerating/decelerating phase of
expansion. The negatively/positively evolving deceleration parameter
discovers accelerating/decelerating phases of expanding cosmos while
constant expansion appears for $q=0$. The effective EoS parameter
classifies these accelerating/decelerating cosmic phases into
distinct regimes as radiation ($\omega_{eff}=\frac{1}{3}$), matter
($\omega_{eff}=0$), quintessence ($-1<\omega\leq-1/3$) and phantom
DE ($\omega<-1$) regimes. The mathematical forms of these standard
parameters are given as
\begin{equation}\nonumber
H=\frac{\dot{a}}{a},\quad q=-\frac{\dot{H}}{H^2}-1,\quad
w_{eff}=\frac{p_{_{eff}}}{\rho_{_{eff}}}=\frac{p_m+p_d}{\rho_m+\rho_d}.
\end{equation}
For graphical interpretation, we choose Mathemtica software with
particular values of arbitrary constants that are mentioned in the
caption of figures. The upper panel of figure \textbf{1} refers to
accelerated expansion as both scale factor and Hubble parameter are
positively increasing. In figure \textbf{1}, the lower panel
interprets evolution of deceleration (left) and effective EoS
(right) parameters. Initially, both parameters correspond to
accelerated expanding cosmos whereas a phase transition from
accelerated to decelerated universe is observed with the passage of
time. The deceleration parameter specifies constant state of
expansion for $q=0$ while the effective EoS parameter identifies
radiation-dominated phase of the universe.
\begin{figure}\epsfig{file=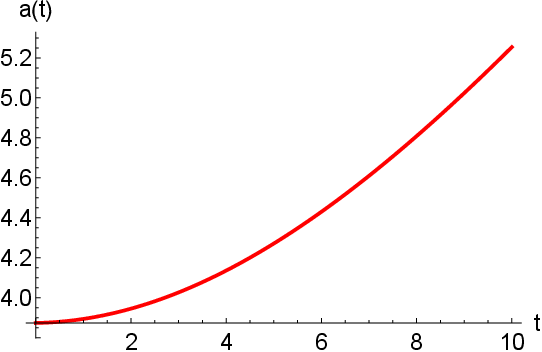,
width=0.45\linewidth}\epsfig{file=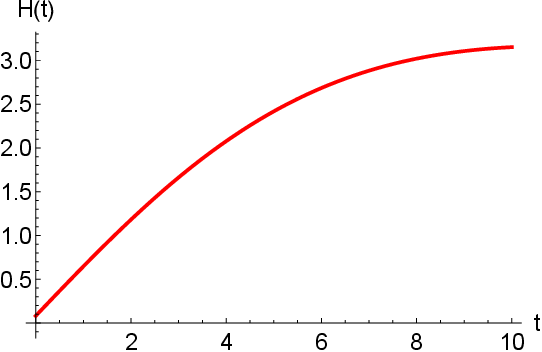,
width=0.45\linewidth}\\
\epsfig{file=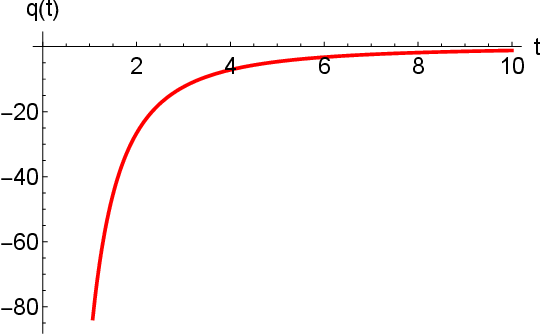,width=0.45\linewidth}
\epsfig{file=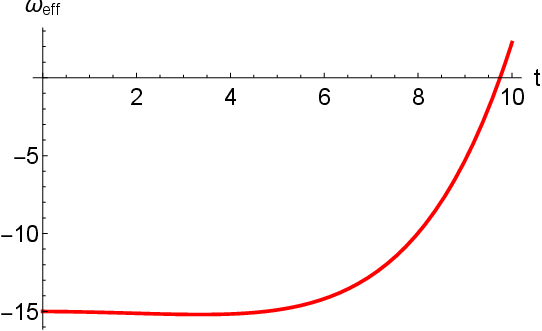,width=0.45\linewidth}\caption{Plots of
cosmological parameters versus cosmic time $t$ for $c_2=2$,
$c_3=c_4=0.5$, $c_{9}=15$, $c_{11}=0.35$ and $c_{12}=0.05$.}
\end{figure}

Among gravitational theories, the $f(R)$ theory explores cosmic
evolution as well as current expansion in the most remarkable
manner. The formalism of this theory puts forward such spectacular
models with positive/negative powers of curvature terms that unravel
enigmatic mysteries behind early as well as late-time universe
\cite{b5}. Besides this revolutionary incentive, the negative
curvature terms may induce unfeasible behavior. This problem can be
sorted out by imposing some constraints on higher-order curvature
derivatives, i.e., $f_R>0,~f_{_{RR}}>0$ with $R>R_0$ ($R_0$ defines
current scalar curvature) \cite{b6}. In non-minimally coupled
gravitational theories, the appearance of an extra force also
introduces instabilities against local perturbations \cite{b1a}.
These instabilities can be avoided by using Dolgov-Kawasaki
instability criteria that demand positivity of higher-order
curvature terms and matter variables \cite{m1}. In case of
non-minimally coupled $f(R,T)$ theory, an additional constraint is
suggested such as $1+f(R,T)_{T}>0$ \cite{b1}. For $f(R,T,Q)$
gravity, the Dolgov-Kawasaki instability analysis introduces two
more conditions to achieve viable behavior given by
\begin{equation}\label{a}
\frac{1+f_{_{T}}+1/2Rf_{_{Q}}}{f_{_{R}}-\mathcal{L}_mf_{_{Q}}}>0,
\quad 3f_{_{RR}}+(1/2T-T^{00}f_{_{RQ}})\geq0.
\end{equation}

In order to determine stable/unstable nature of constructed models,
the squared speed of sound comes up with simple criteria. The
established model preserves its stable/unstable state against
background perturbations for positively/negatively evolving squared
speed of sound. The $r-s$ parameters measure compatibility between
constructed and well-known cosmological models for their particular
values. A model is said to be consistent with $\Lambda$CDM, CDM
models and Einstein universe if ($r,s$)=(1,0), (1,0) and
(-$\infty$,$\infty$), respectively. For $s>0$ with $r<1$ and $s<0$
with $r>1$, the established model can be characterized as
quintessence, phantom DE and Chaplygin gas models, respectively. The
squared speed of sound and $r-s$ parameters are given as
\begin{equation*}
v_s^2=\frac{\dot{p_d}}{\dot{\rho_d}},\quad
r=q+2q^2-\frac{\dot{q}}{H},\quad s=\frac{r-1}{3(q-\frac{1}{2})}.
\end{equation*}

In the distinct regimes of expanding cosmos, the evaluation of
fractional densities provides a better understanding of cosmological
structure and matter distribution. According to Planck observational
data, the fractional densities of flat cosmos are restricted to
follow $\Omega_m+\Omega_d=1$, where $\Omega_d\simeq0.6889$ and
$\Omega_m\simeq0.3111$ \cite{b8}. Mathematically, these fractional
densities are given by
\begin{equation*}
\Omega_m=\frac{\rho_m}{3m_{pl}^2H^2},\quad
\Omega_d=\frac{\rho_d}{3m_{pl}^2H^2}.
\end{equation*}
Figure \textbf{2} determines stable as well as viable $f(R,Q)$ model
due to positive nature of the squared speed of sound and viability
constraints. In figure \textbf{3} (left plot), the $r-s$ parameters
measure a consistency of established $f(R,Q)$ model with
quintessence and phantom regions as $s>0$ when $r<1$. The right plot
of figure \textbf{3} shows that initially, the fractional densities
follow Planck observational constraint while this consistency is
disturbed due to the dominance of $\Omega_m$ over $\Omega_d$ as time
grows.
\begin{figure}\epsfig{file=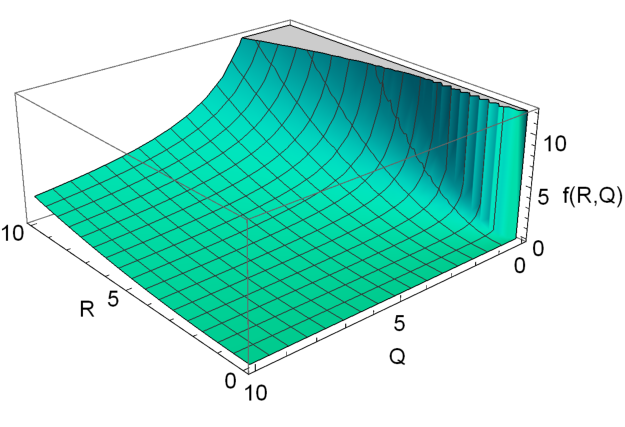,
width=0.45\linewidth}\epsfig{file=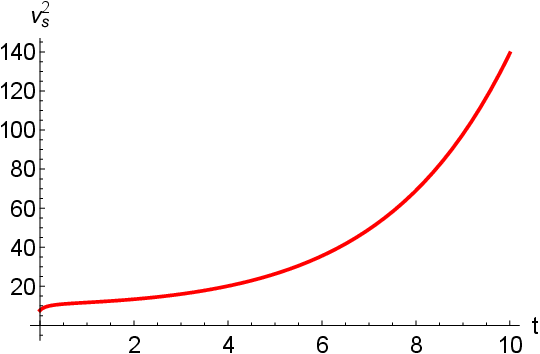, width=0.45\linewidth}\\
\epsfig{file=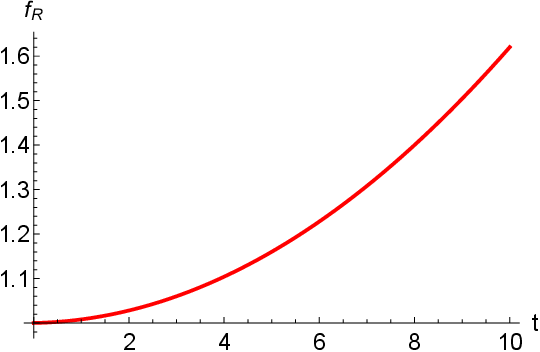, width=0.45\linewidth}\epsfig{file=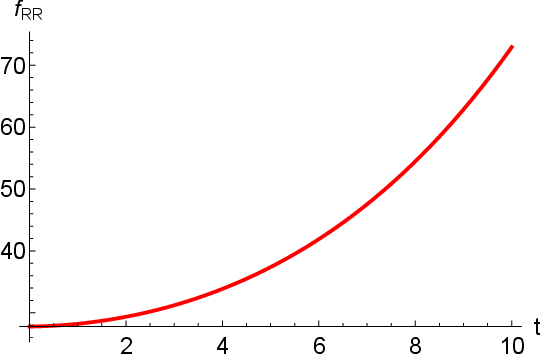,
width=0.45\linewidth}\\
\epsfig{file=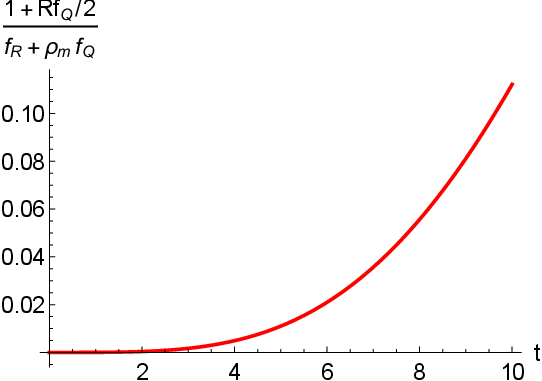, width=0.45\linewidth}\epsfig{file=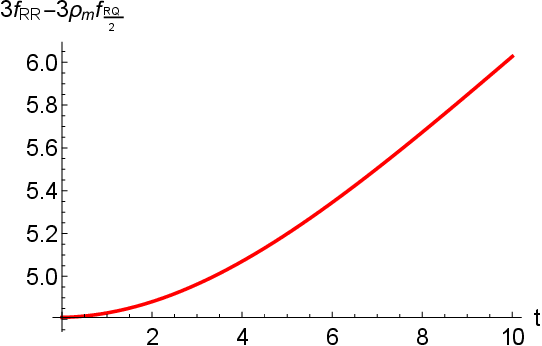,
width=0.45\linewidth}\caption{Plots of constructed $f(R,Q)$ model,
squared speed of sound and viability conditions versus cosmic time
$t$.}
\end{figure}
\begin{figure}\epsfig{file=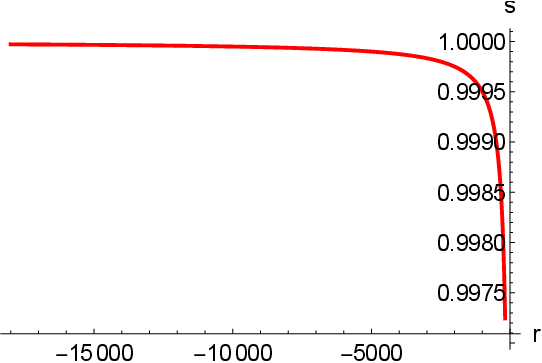,
width=0.45\linewidth}\epsfig{file=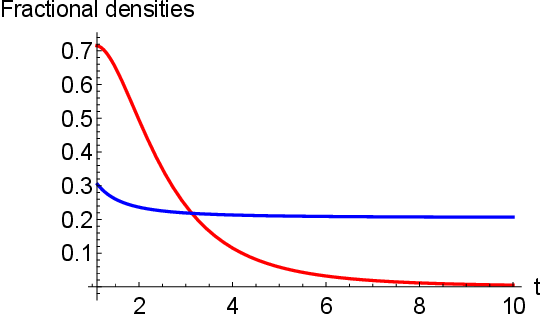,
width=0.45\linewidth}\caption{Plots of $r-s$ parameters (left) and
fractional densities (right) $\Omega_{m}$ (blue) and $\Omega_{d}$
(red) versus cosmic time $t$.}
\end{figure}

\section{Noether Symmetries of \textbf{$f(R,T)$ Model}}

In this case, we study the impact of non-minimal curvature-matter
coupling in the background of pressureless fluid while the function
$f$ is considered to be independent of the term $Q$. For these
restrictions, we formulate all possible Noether symmetries and their
associated conserved entities in the absence of boundary term
($\mathcal{B}=0$). For this purpose, the solutions to over
determining system of non-linear equations (\ref{20a})-(\ref{39a})
are given as
\begin{eqnarray}\nonumber
\alpha&=&c_6,\quad\beta=\frac{c_1c_5(c_3R-2c_2T^2)+c_1^2c_3a^3R-2c_1c_2a^3T^2
(c_1+T)}{c_3^2a(c_5+c_1a^3+a^3T)},\quad\psi=0,\\\label{46a}\phi&=&\frac{c_1T}{c_3a},
\quad\rho_m(a)=c_1+\frac{c_5}{a^3},\quad f(R,T)=(c_3\ln
T+c_4)R+c_5T^2+c_7,
\end{eqnarray}
where $c_i's~(i=1,2,...,7$) represent constants. In the case of
simple Noether symmetry, we obtain $f(R,T)$ model that incorporates
a direct non-minimal coupling between curvature and matter
variables. For the sake of simplicity, we redefine constants as
$c_1c_3c_5=d_1,~c_1^2c_3=d_2,~c_1c_2c_5=d_3,
~c_1^2c_2=d_4,~c_1c_2=d_5,~c_1/c_3=d_6,~c_3^2c_5=d_7,c_1c_3^2=d_8$
and $c_3^2=d_9$. For these constants, the group of Noether
symmetries together with conserved integrals are
\begin{eqnarray}\label{48a}
Y_1&=&\partial_{_a},\quad
I_1=12a\dot{a}f_{_{R}}+6a^2(\dot{R}f_{_{RR}}+\dot{T}f_{_{RQ}}),\\\nonumber
Y_2&=&\frac{R}{d_7a+a^4(d_8+d_9T)}\partial_{_{R}},\quad
I_2=\frac{6a^2\dot{a}f_{_{RR}}R}{d_7a+a^4(d_8+d_9T)},\\\nonumber
Y_3&=&\frac{T}{a}\partial_{_{T}},\quad I_3=6a\dot{a}f_{_{RT}}T,
\\\nonumber
Y_4&=&\frac{a^3R}{d_7a+a^4(d_8+d_9T)}\partial_{_{R}},\quad
I_4=\frac{6a^5\dot{a}f_{_{RR}}R}{d_7a+a^4(d_8+d_9T)},\\\nonumber
Y_5&=&-\frac{2T^2}{d_7a+a^4(d_8+d_9T)}\partial_{_{R}},\quad
I_5=-\frac{12a^2T^2\dot{a}f_{_{RR}}}{d_7a+a^4(d_8+d_9T)},\\\nonumber
Y_6&=&-\frac{2a^3T^2}{d_7a+a^4(d_8+d_9T)}\partial_{_{R}},\quad
I_6=-\frac{12a^5T^2f_{_{RR}}}{d_7a+a^4(d_8+d_9T)},\\\nonumber
Y_7&=&-\frac{2a^3T^3}{d_7a+a^4(d_8+d_9T)}\partial_{_{R}},\quad
I_7=-\frac{12a^5T^3f_{_{RR}}}{d_7a+a^4(d_8+d_9T)}.
\end{eqnarray}
The correspondence of standard symmetries and conserved quantities
cannot be established with the above set of seven symmetry
generators and conserved integrals in the absence of boundary term.
In order to study cosmological behavior of non-minimally coupled
$f(R,T)$ model (\ref{46a}), we formulate exact solution of the scale
factor from Eq.(\ref{48a}) given by
\begin{eqnarray}\nonumber
a(t)&=&\frac{6I_1c_5}{t+c_{8}}\left[c_3Rootof(\mathcal{K})
e^{-Rootof(\mathcal{K})}\right],
\end{eqnarray}
where $\mathcal{K}=I_1^3(t+c_{8})e^{2\mathcal{Z}}
-216c_5^2(c_3\mathcal{Z}+c_4)^3$ and $c_{8}$ indicates integration
constant whereas $\mathcal{Z}$ denotes global variable.
\begin{figure}\center{\epsfig{file=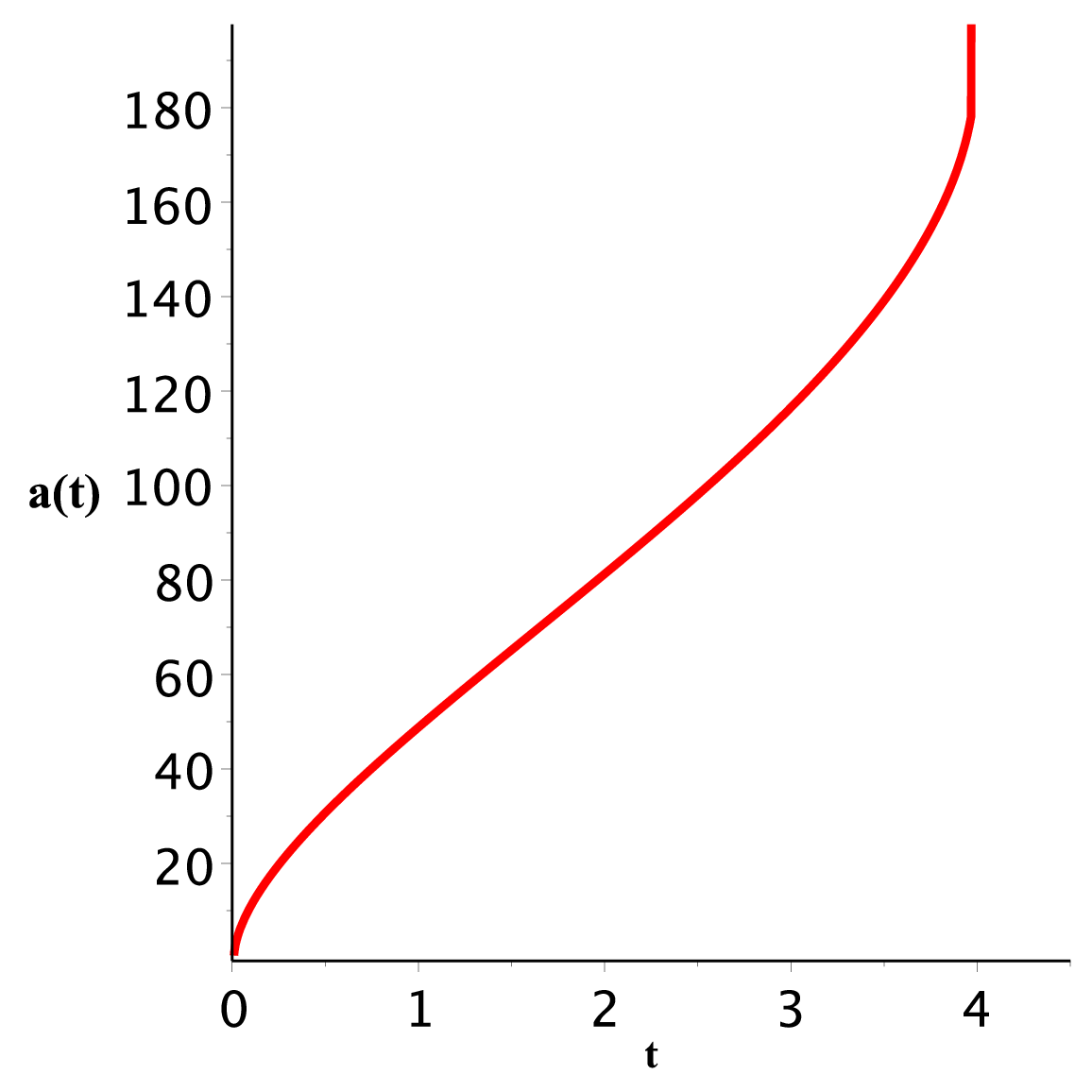,
width=0.4\linewidth}\epsfig{file=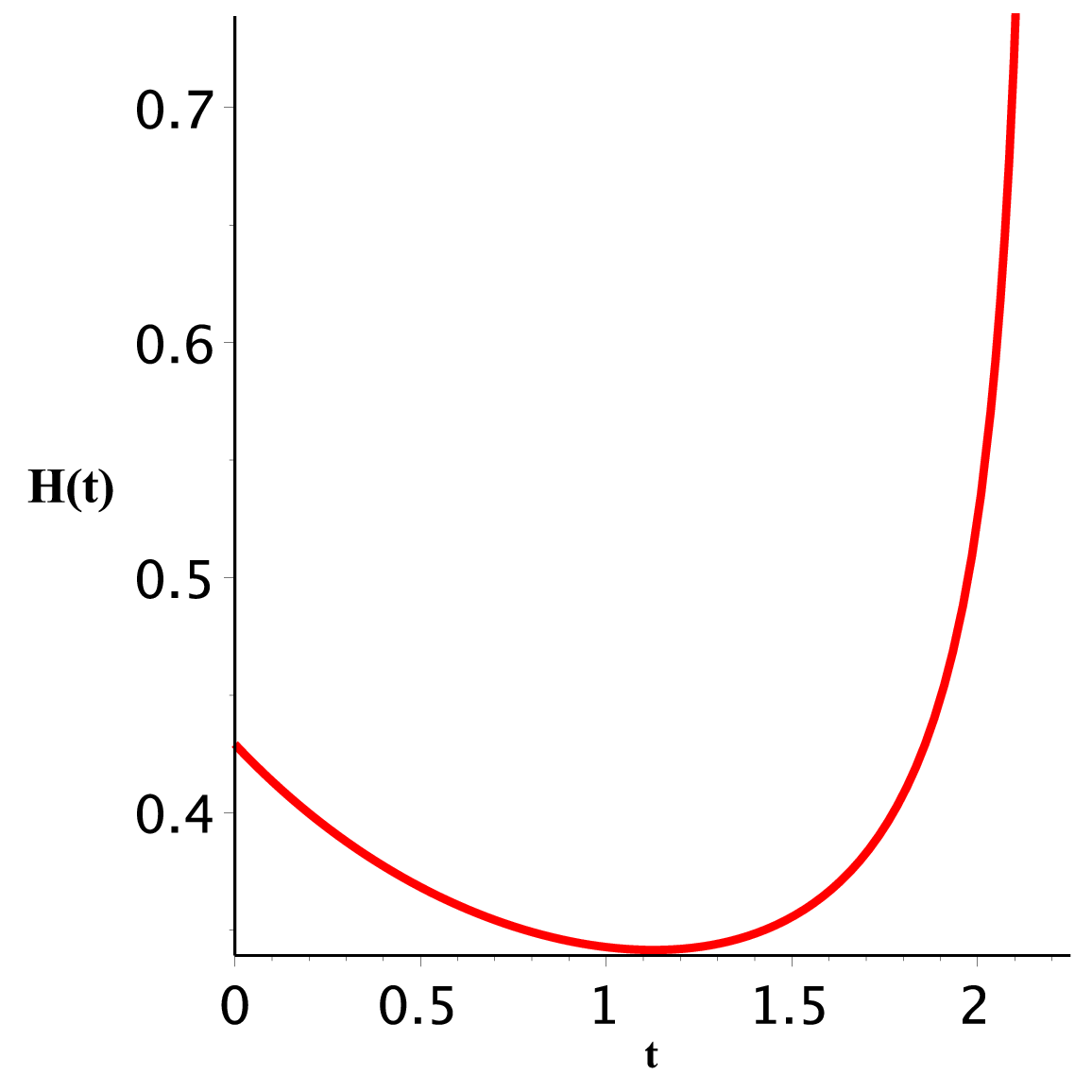,
width=0.4\linewidth}\\
\epsfig{file=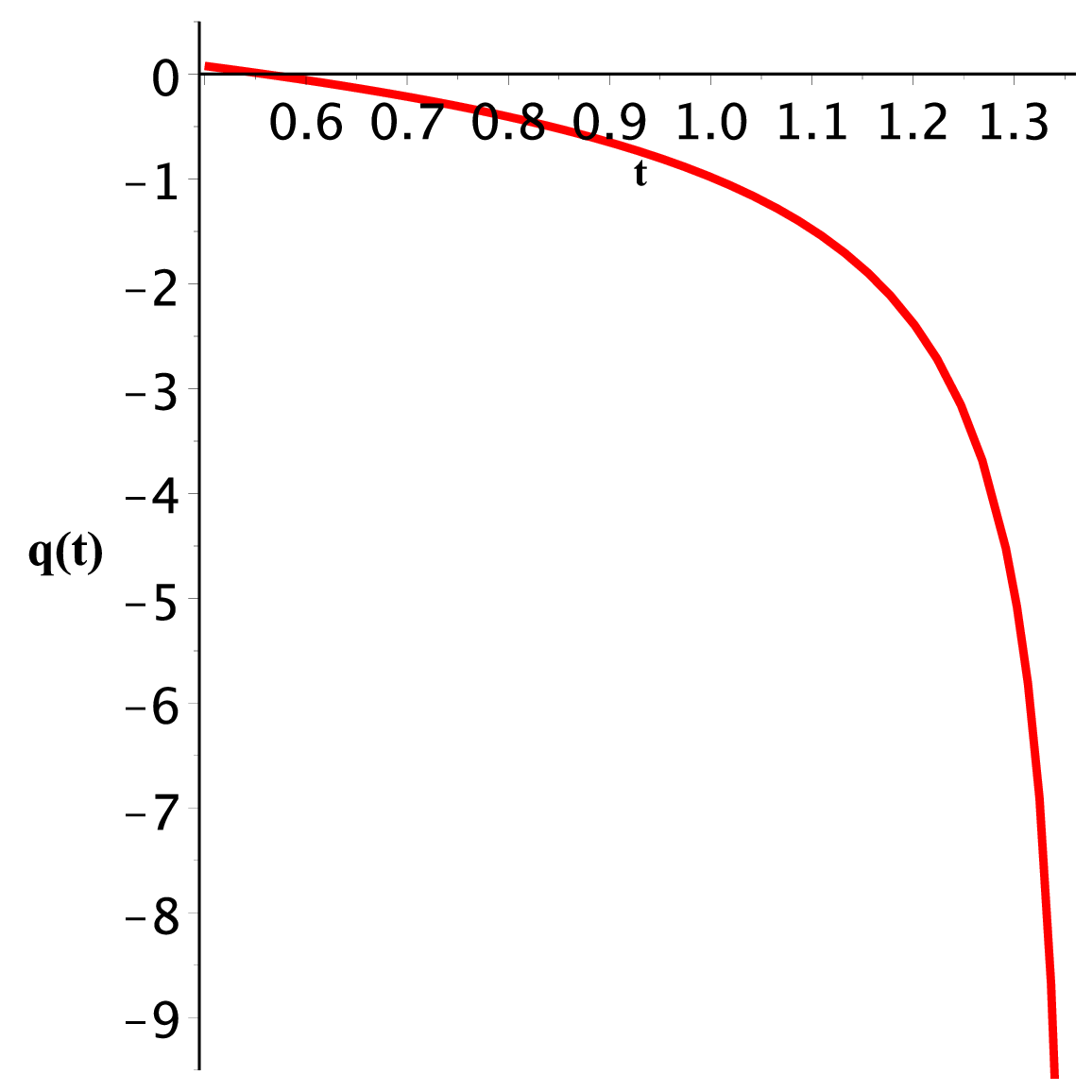,width=0.4\linewidth}
\epsfig{file=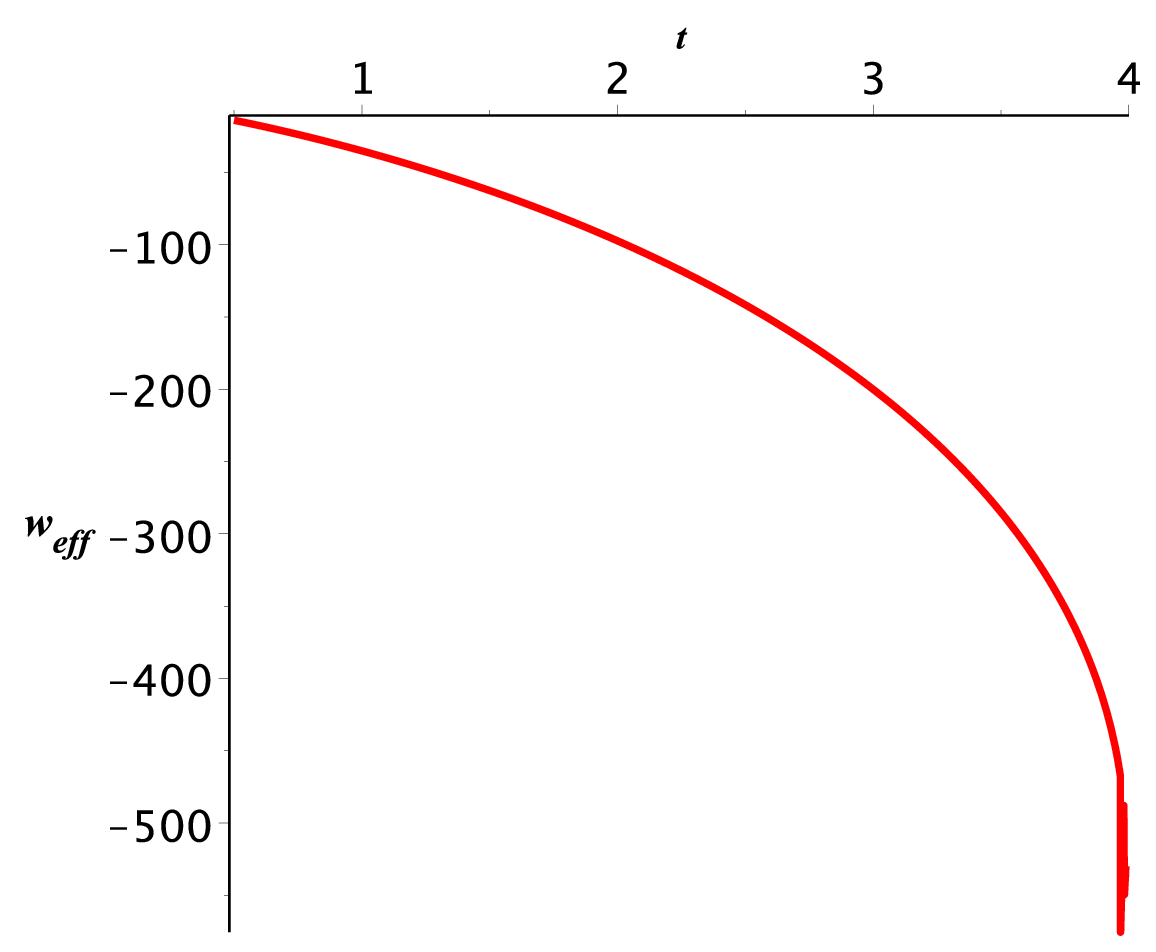,width=0.4\linewidth}}\caption{Plots of
cosmological parameters versus cosmic time $t$ for $c_1=-0.01$,
$I_1=c_4=10$, $c_5=c_8=0.01$, $c_{3}=1.25$, $c_{7}=2.5$ and
$c_{12}=0.05$.}
\end{figure}
\begin{figure}\center{\epsfig{file=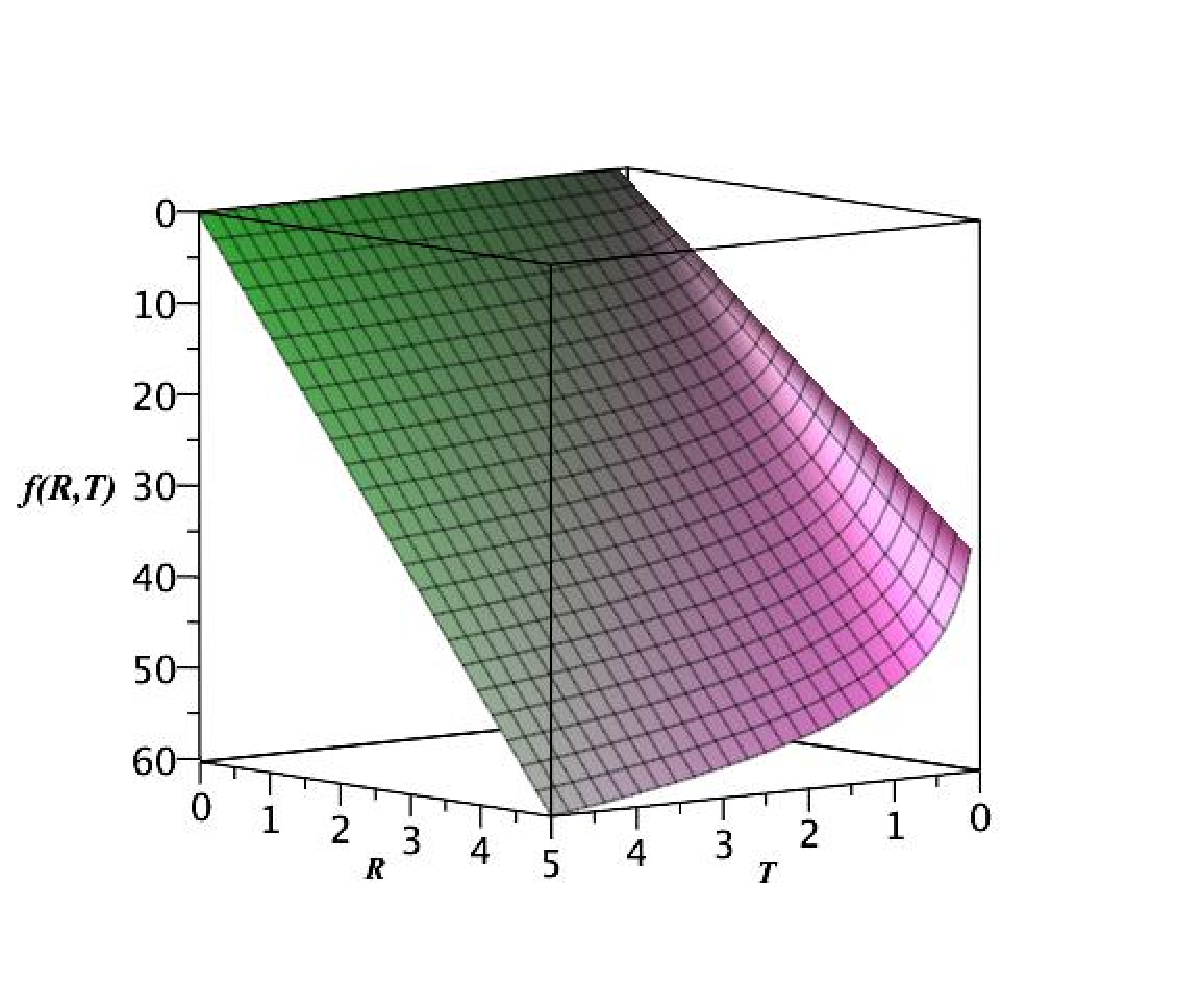,
width=0.45\linewidth}\epsfig{file=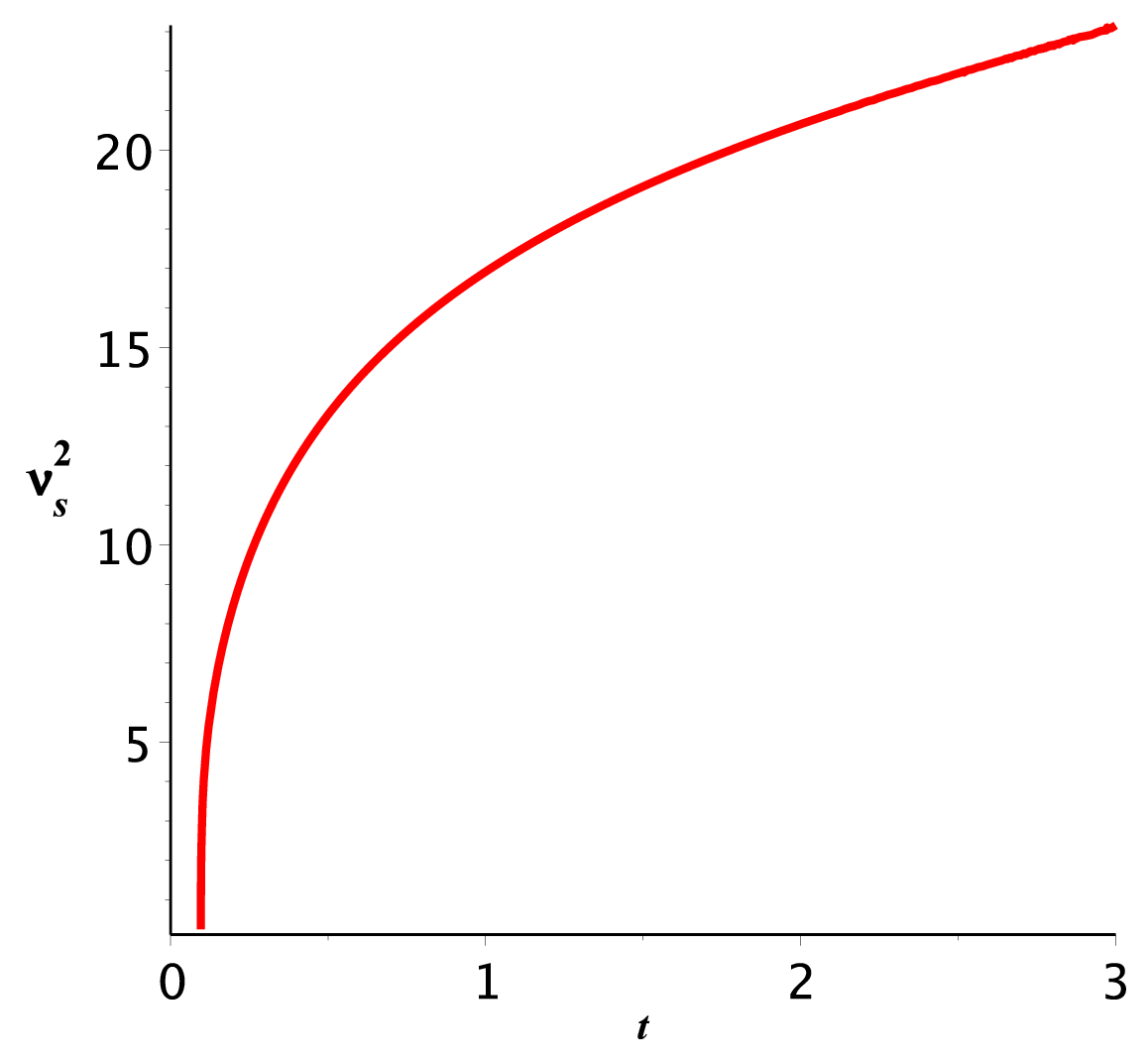, width=0.38\linewidth}\\
\epsfig{file=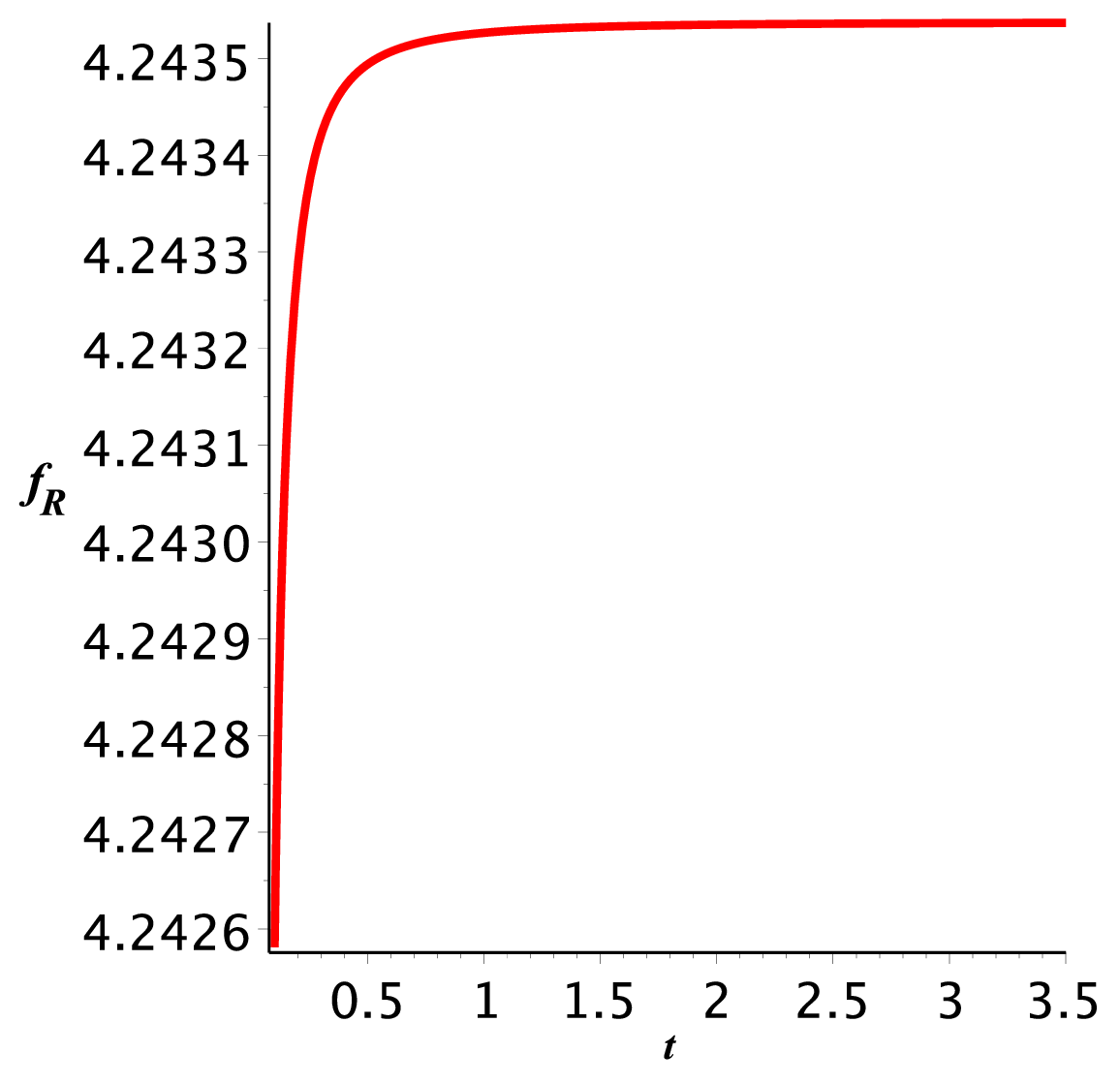, width=0.38\linewidth}\epsfig{file=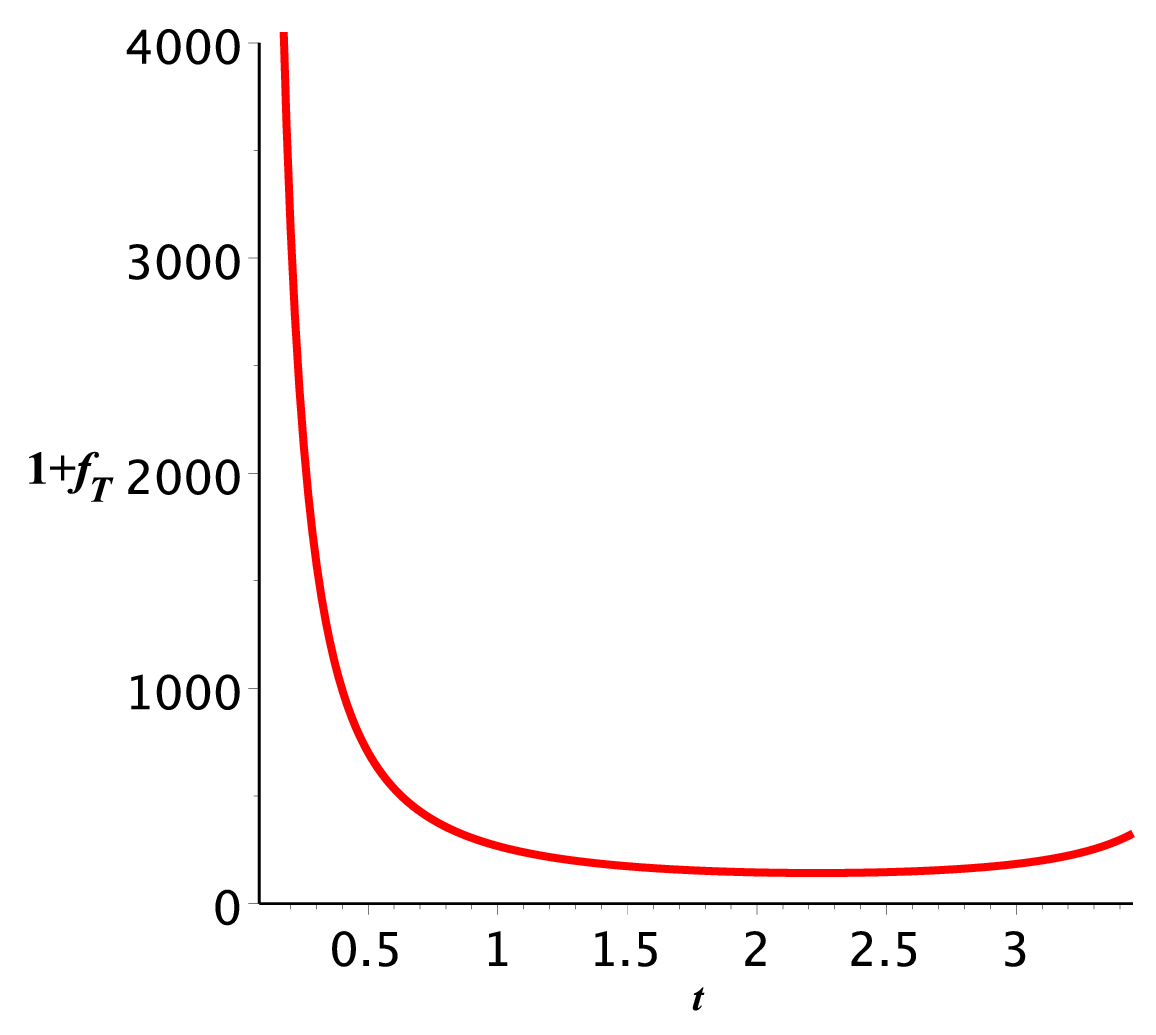,
width=0.38\linewidth}}\caption{Plots of constructed $f(R,T)$ model,
squared speed of sound and viability conditions versus cosmic time
$t$.}
\end{figure}
\begin{figure}\center{\epsfig{file=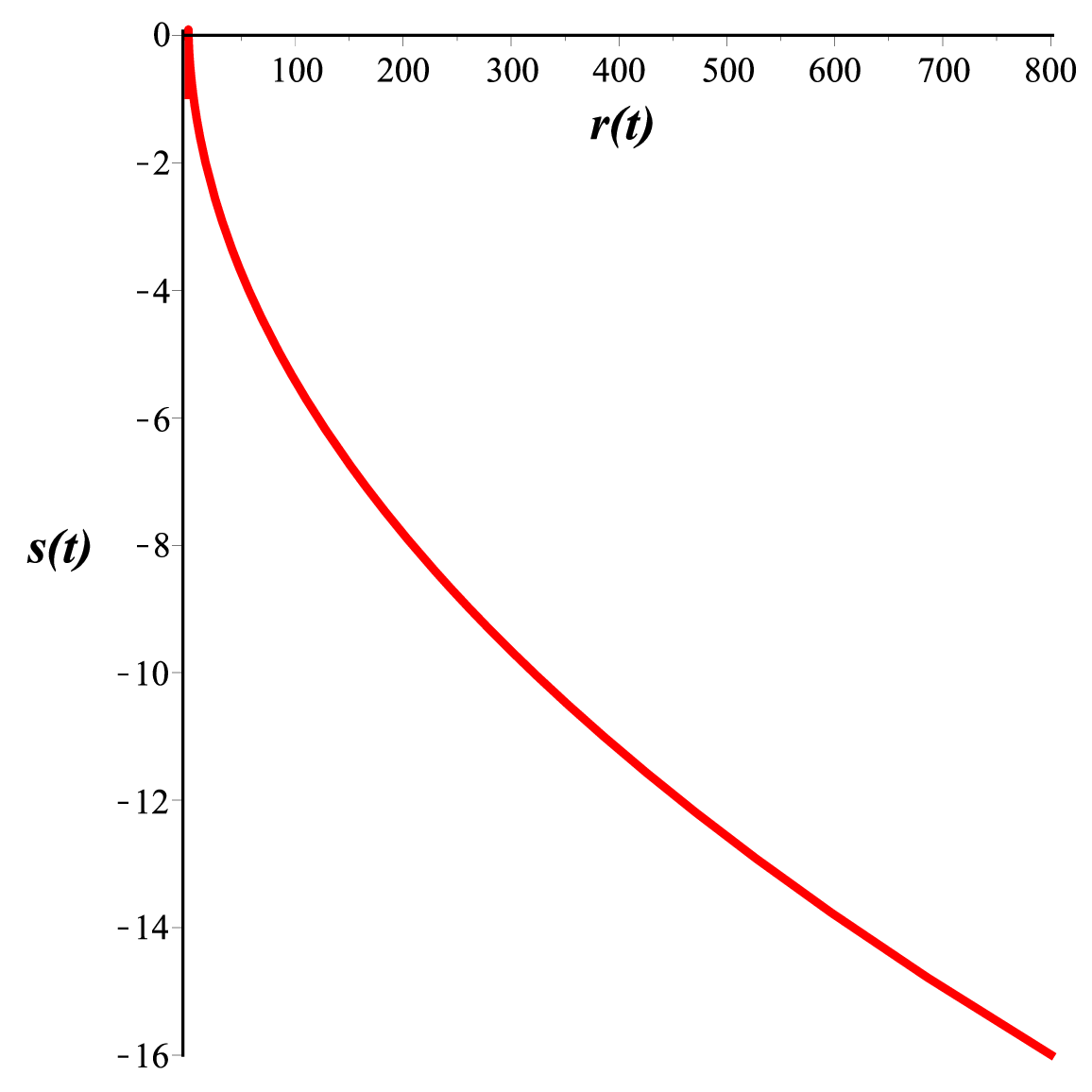,
width=0.42\linewidth}\epsfig{file=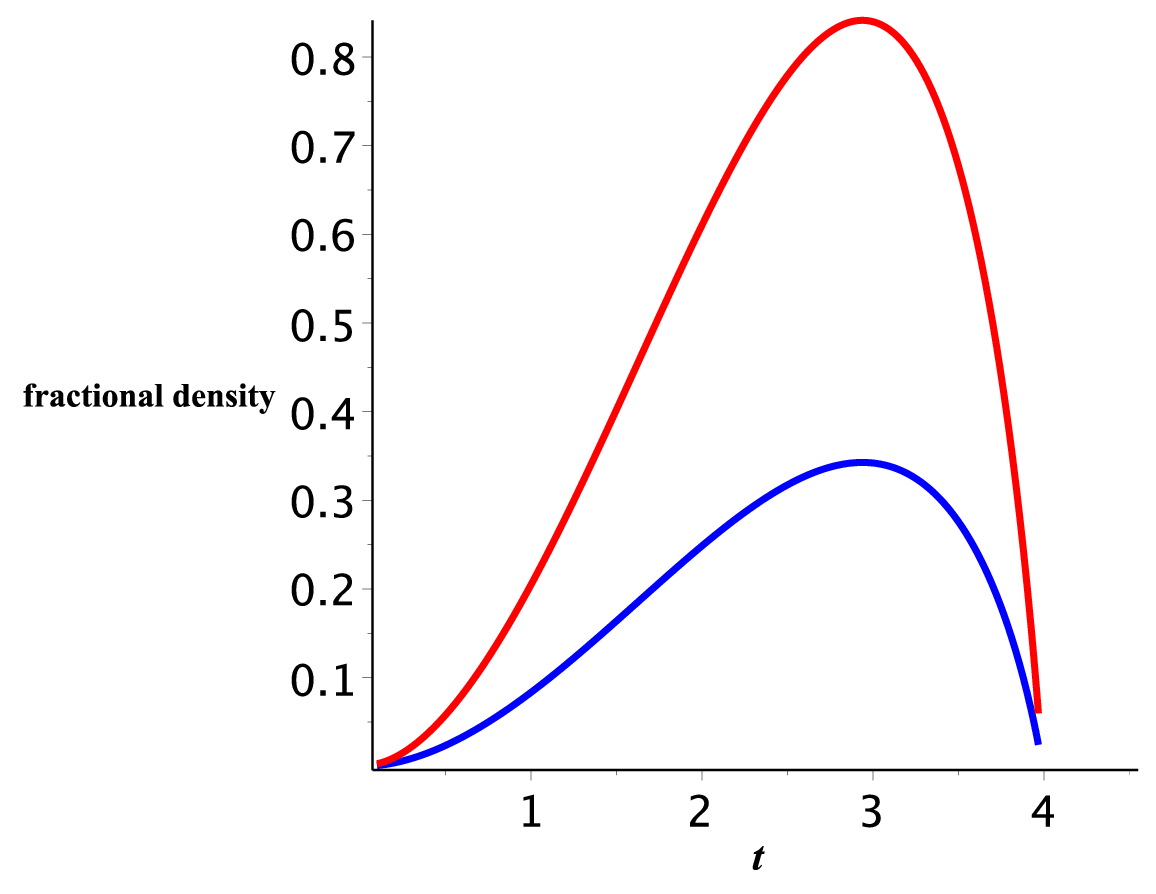,
width=0.5\linewidth}}\caption{Plots of $r-s$ parameters (left) and
fractional densities (right) $\Omega_{m}$ (blue) and $\Omega_{d}$
(red) versus cosmic time $t$.}
\end{figure}

Figure \textbf{4} develops cosmological understanding for exact
solution of scale factor through Hubble, deceleration and effective
EoS parameters. The graphical representation of these parameters
ensures accelerated cosmic expansion while EoS further characterizes
a smooth transition from quintessence to phantom DE era. Figure
\textbf{5} measures the evolution as well as realistic nature of
non-minimally coupled $f(R,T)$ model via squared speed of sound and
viability constraints. It is interesting to mention here that the
established model is stable and viable in the background of
accelerated expanding cosmos. In Figure \textbf{6}, the model is
found to be consistent with Chaplygin gas model as $r>1$ and $s<0$
(left plot) while the dominance of DE energy density over matter
energy density shows consistency with recent observational data
(right plot).

\section{Noether Symmetries of \textbf{$f(T,Q)$ Model}}

In this section, the source of non-minimal interactions is
considered to be trace of energy-momentum tensor, contracted Ricci
and energy-momentum tensors while the generic function is considered
to be free from curvature invariant. To investigate the effect of
these interactions, we solve the system (\ref{20a})-(\ref{39a}) to
determine possible symmetries and corresponding conservation laws
for $f(T,Q)$ model. In this regard, we consider
$\phi(a,R,T,Q),_R=\psi(a,R,T,Q)_R=0$ and obtain the resulting
solution as follows
\begin{eqnarray}\nonumber
\alpha&=&c_1,\quad\beta=0,\quad\phi=-\frac{c_6Q}{c_3a^2(c_7+T)},
\quad\psi=\frac{c_8}{c_{9}a^2},\\\label{46b}f(T,Q)&=&c_3T+c_4Q+c_5,
\quad\rho_m(a)=c_1a+c_2.
\end{eqnarray}
The formulated $f(T,Q)$ model admits implicit non-minimal
interactions while the generators and respective conserved integrals
of simple Noether symmetry turn out to be
\begin{eqnarray}\label{48b}
Y_1&=&\partial_{_a},\quad
I_1=-6a\dot{a}f_{_{Q}}(2\rho_m+a\rho_m')+3a^2\rho_m
(\dot{T}f_{_{TQ}}+\dot{Q}f_{_{QQ}}),\\\nonumber
Y_2&=&-\frac{Q}{a^2c_3(c_7+T)}\partial_{_{T}},\quad
I_2=\frac{3\dot{a}\rho_mQf_{_{TQ}}}{c_3(c_7+T)},\\\nonumber
Y_3&=&\frac{1}{a^2c_9}\partial_{_{Q}},\quad
I_3=-\frac{3\dot{a}\rho_mf_{_{QQ}}}{c_9}.
\end{eqnarray}
The symmetry generator $Y_1$ defines scaling symmetry and
correspondingly, $I_1$ refers to conservation of linear-momentum.
Besides the conserved linear-momentum, the first Noether integral
(\ref{48b}) also helps to determine exact solution for $I_1=0$ given
as
\begin{eqnarray*}
a(t)&=&\left[-LambertW\left(-\frac{e^{\frac{2c_1c_{10}+2tc_1-c_2}
{c_2}}}{c_2}\right)+\frac{2c_1c_{10}+2tc_1-c_2}{c_2}\right]
\\\nonumber&\times&\frac{2c_1c_{10}+2tc_1-c_2}{c_1}.
\end{eqnarray*}

In Figure \textbf{7}, the scale factor as well as cosmological
parameters measure an accelerated expansion that favors a smooth
transition from quintessence to phantom DE phase. The positively
evolving minimally coupled $f(T,Q)$ model is found to be unstable at
early stage as $v_s^2<0$ while the model attains stability as time
passes (Figure \textbf{8}, upper panel). It is interesting to
mention here that the established model satisfies all viability
constraints for $c_3,~c_4\geq0$ whereas $r-s$ parameters specify
correspondence with Chaplygin gas model (Figure \textbf{8}, lower
panel). Figure \textbf{9} indicates that at $t=0.34$, the fractional
densities relative to matter and DE are found to be compatible with
Planck's observational data.
\begin{figure}\center{\epsfig{file=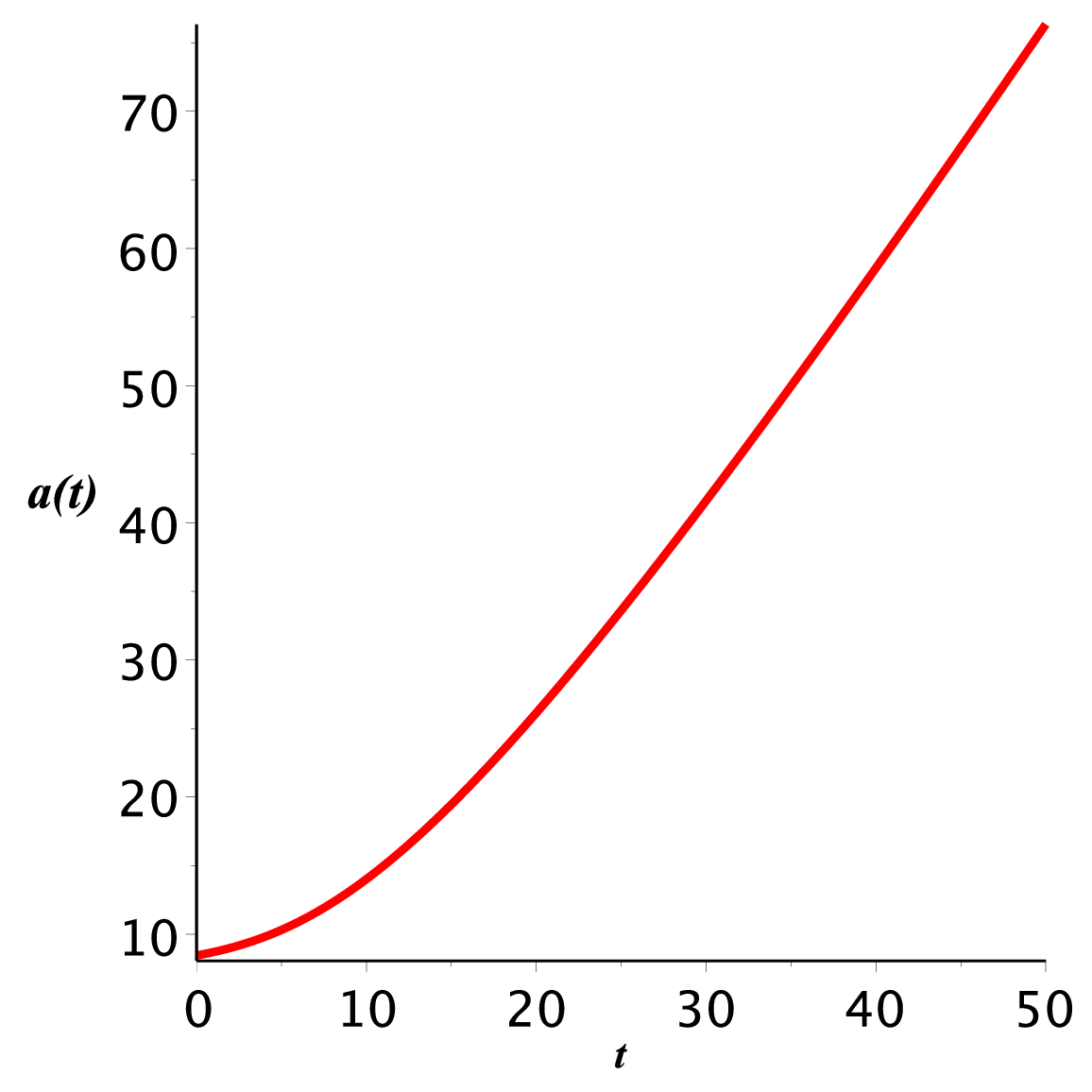,
width=0.35\linewidth}\epsfig{file=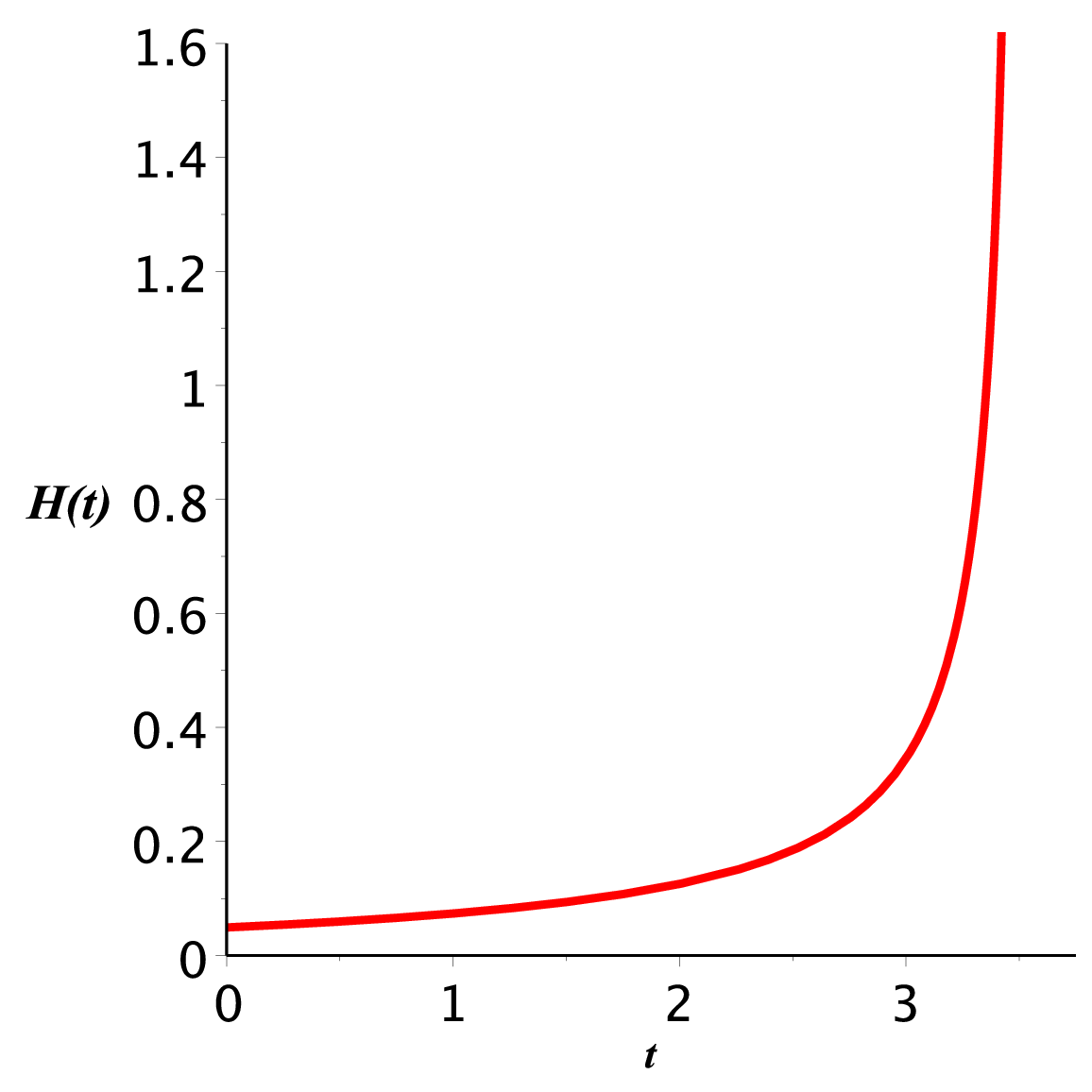,
width=0.35\linewidth}\\
\epsfig{file=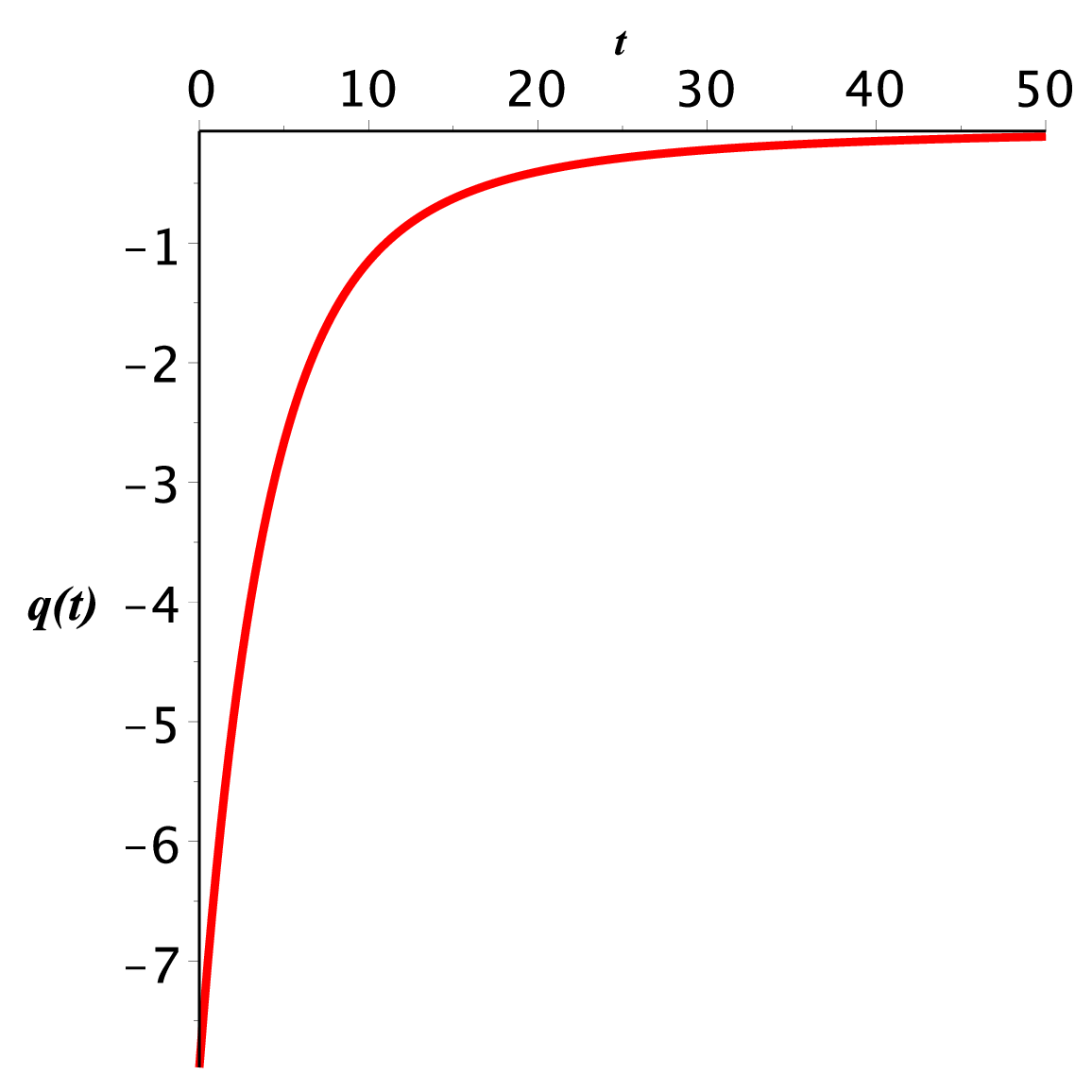,width=0.35\linewidth}
\epsfig{file=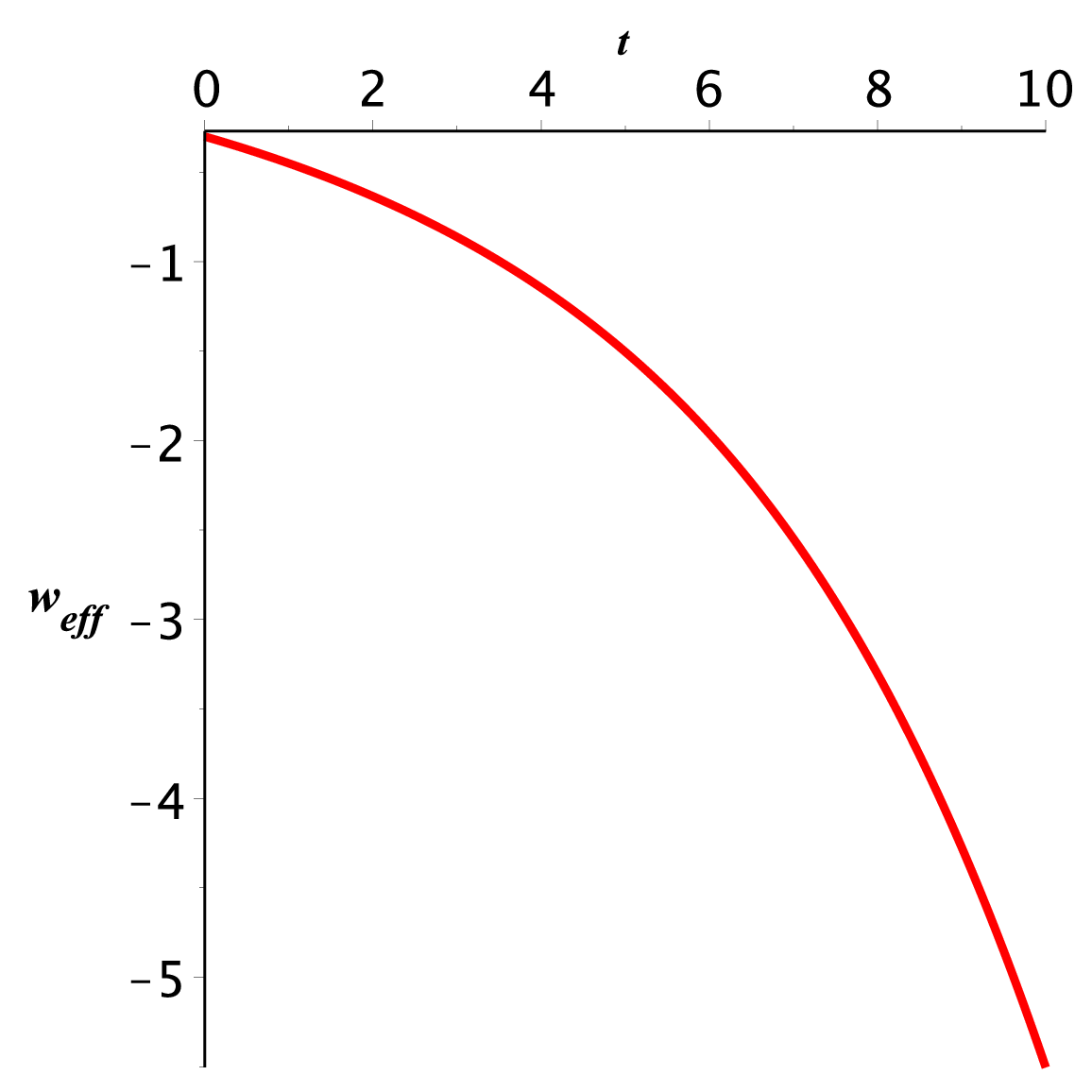,width=0.35\linewidth}}\caption{Plots of scale
factor, Hubble, deceleration and effective EoS parameters versus
cosmic time $t$ for $c_1=-0.2$, $c_2=-1.5$, $c_3=2$, $c_4=1.05$,
$I_{2}=0$, $c_{5}=1$ and $c_{10}=-2$.}
\end{figure}
\begin{figure}\center{\epsfig{file=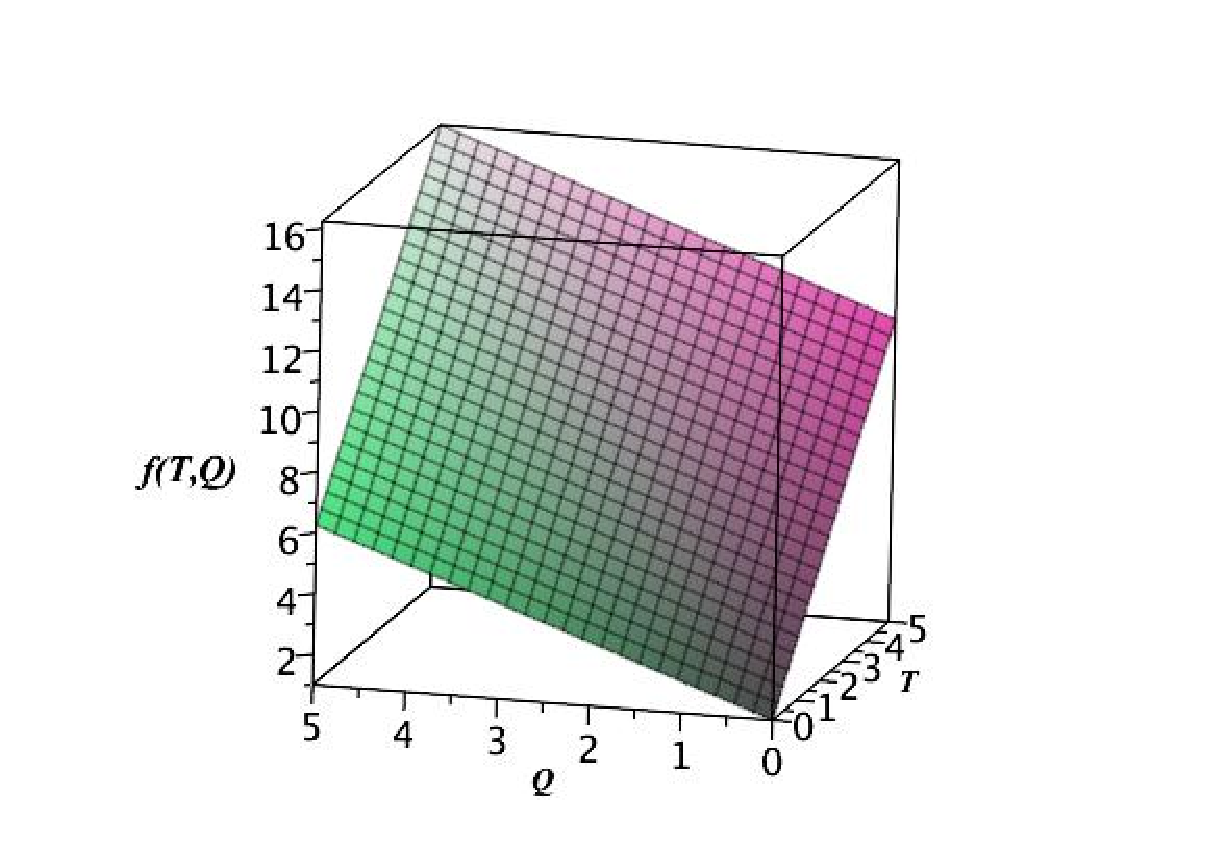,
width=0.6\linewidth}\epsfig{file=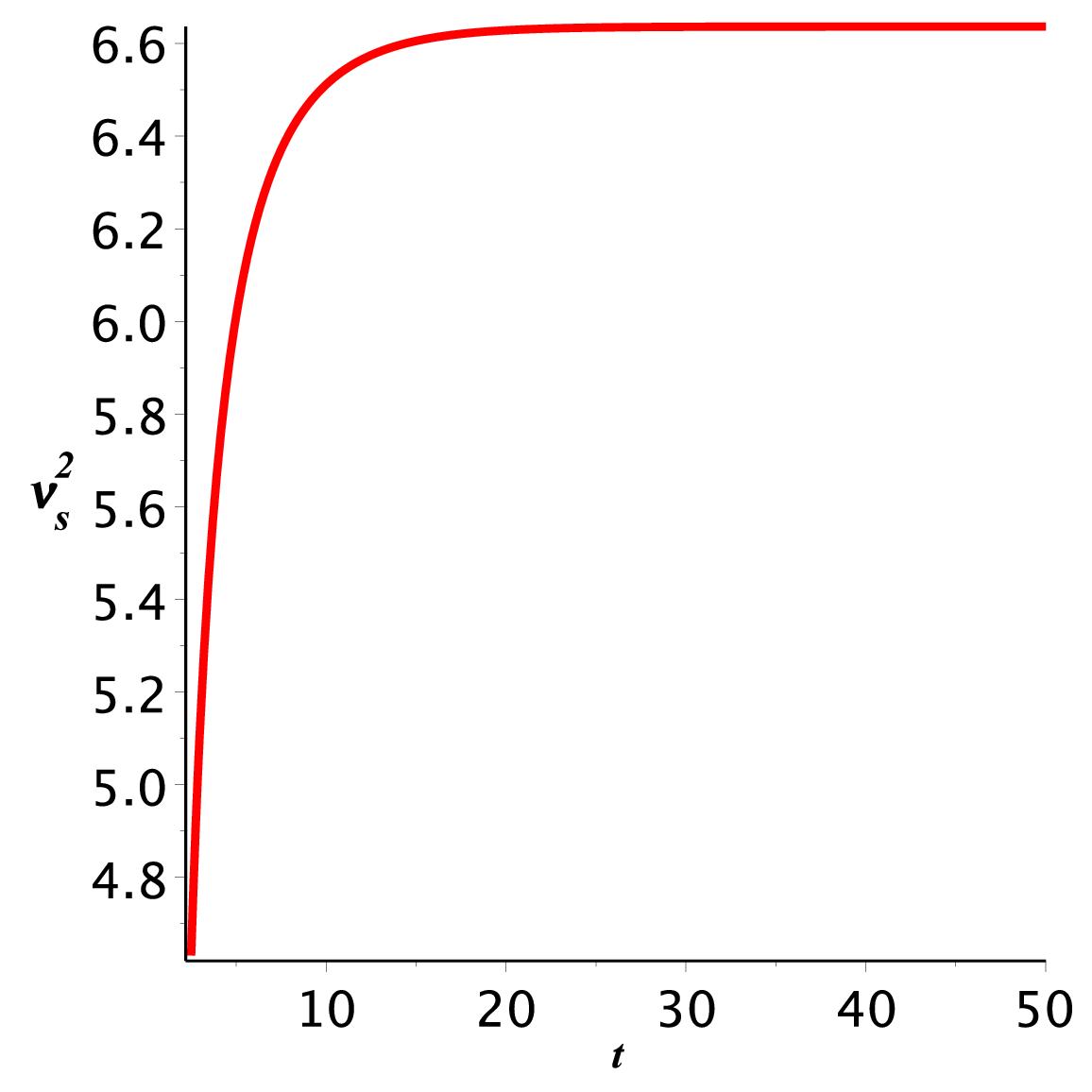,
width=0.35\linewidth}\\
\epsfig{file=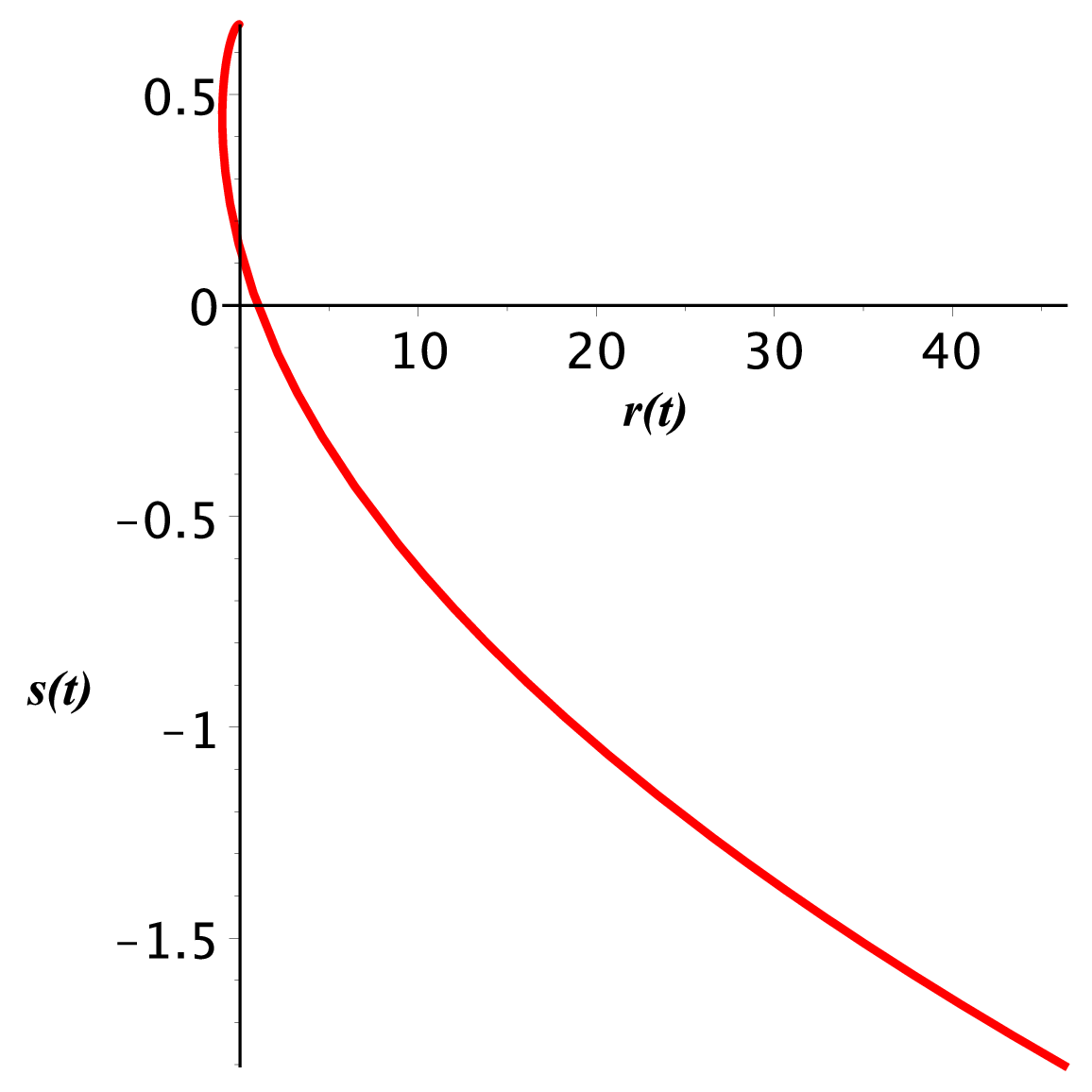, width=0.35\linewidth}}\caption{Plots of
$f(T,Q)$ (upper left), squared speed of sound (upper right),
viability condition (lower left) and $r-s$ parameters (lower right)
versus cosmic time $t$.}
\end{figure}
\begin{figure}\center{\epsfig{file=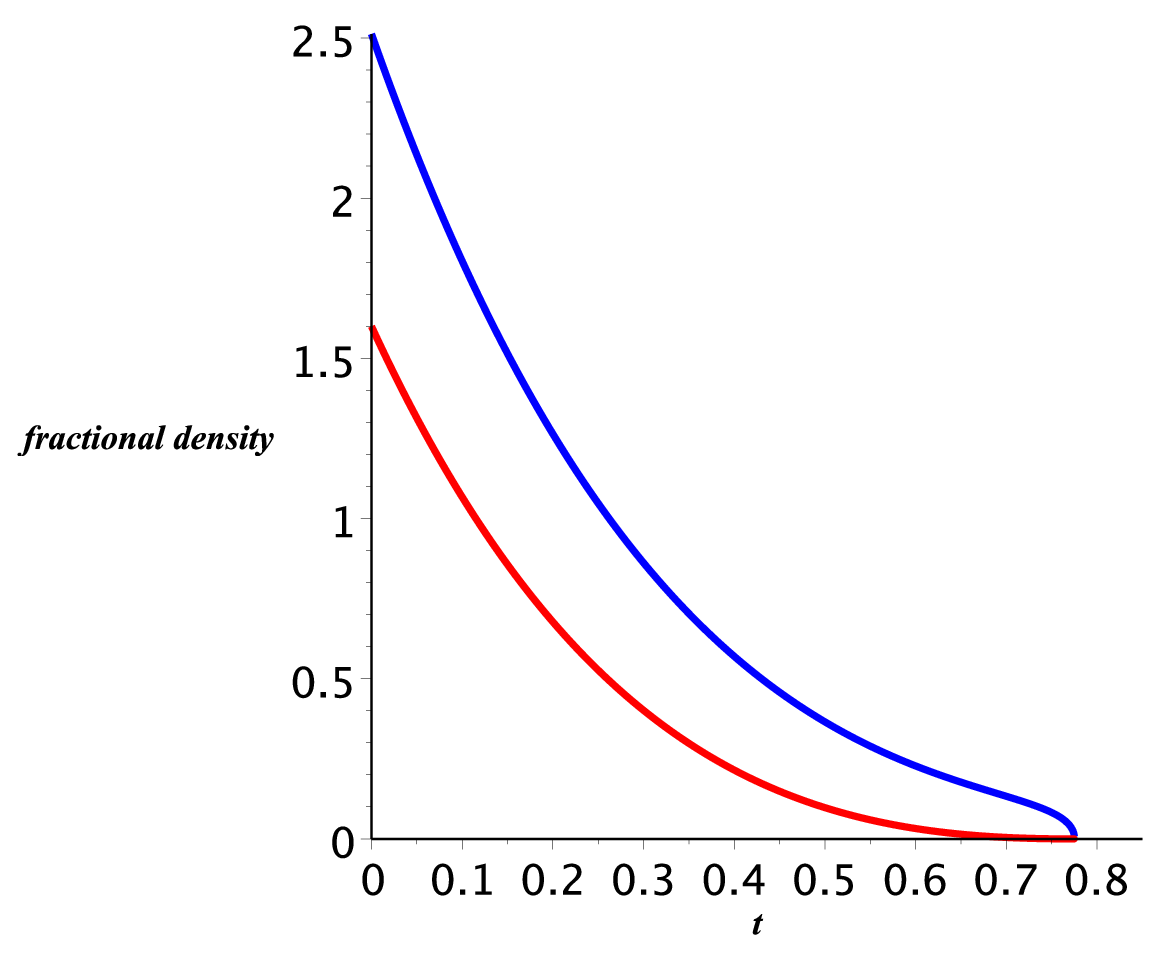,
width=0.45\linewidth}}\caption{Plots of fractional densities (right)
$\Omega_{m}$ (blue) and $\Omega_{d}$ (red) versus cosmic time $t$.}
\end{figure}

\section{Noether Symmetry of \textbf{$f(R,T,Q)$ Model}}

This case determines the existence of Noether point symmetries and
corresponding conserved entities in the presence of non-minimal
scalar curvature and matter variables. In order to solve the system
of non-linear equations (\ref{20a})-(\ref{39a}) for Noether
symmetry, we consider $\alpha=F_1(a)F_2(R)F_3(T)F_4(Q)$ and
$\beta=d_1$, which gives
\begin{eqnarray}\nonumber
&&\alpha=\frac{c_6(Rc_1+c_2)(Tc_4+c_5)(c_3+2a^{\frac{2}{3}}Q)}{c_3a^{\frac{7}{6}}},
\quad\beta=c_8,\quad\phi=-\frac{2a^{\frac{2}{3}}(Tc_4+_5)\psi}{c_4},
\\\nonumber&&\psi=-\frac{6c_6(Tc_4+c_5)(Rc_1+c_2)\Sigma}
{a^{\frac{5}{2}}c_3c_4(c_1a^{\frac{1}{3}}+a T)},
\quad\rho_m(a)=\frac{c_6}{a^{\frac{2}{3}}},
\\\label{44d}&&f(R,T,Q)=c_4T\exp[Rc_1+2a^{\frac{2}{3}}Q+c_3].
\end{eqnarray}
Here, the formulated exponential $f(R,T,Q)$ model admits strong
non-minimal interactions between scalar curvature and matter
variables whereas $\Sigma$ is defined as
\begin{eqnarray*}
\Sigma&=&\frac{(R(Tc_4+c_5)c_1+Tc_3c_4+c_5(c_3-1))Qa^{\frac{4}{3}}}{2}
+Q^2a^2(Tc_4+c_5)+\frac{7c_1c_3c_4}{36}\\\nonumber
&+&\left(\left(\left(\frac{RTc_3}{4}+\frac{7Q}{8}\right)c_4
+\frac{Rc_3c_5}{4}\right)c_1-\frac{c_3c_5}{4}\right)a^{\frac{2}{3}}.
\end{eqnarray*}
The arbitrary constants are denoted by $c_i~(i=1,2,...,6$) while the
resulting solution satisfies over-determining system for
$c_2=c_5=0$. For the sake of simplicity, we redefine a few constants
such as $c_1c_4c_6=d_2,~\frac{c_1c_4c_6}{c_3}=d_3,~c_1^2c_4c_6=d_4$
and $\frac{c_1^2c_4c_6}{c_3}=d_5$. In this case, we find the set of
symmetry generators and respective conserved entities as follows
\begin{eqnarray}\nonumber
Y_1&=&\partial_{_R},\quad
I_1=6a^2\dot{a}f_{_{RR}}-6a^2\dot{a}\rho_mf_{_{RQ}},\\\nonumber
Y_2&=&\frac{2RTQ}{a^{\frac{1}{2}}}\partial_{_{a}}
+\frac{6RT^2Q^2}{c_1a^{\frac{1}{3}}+a T}
\left(2a^{\frac{1}{6}}T\partial_{_{T}}
-\frac{1}{\sqrt{a}}\partial_{_{Q}}\right),\\\nonumber
I_2&=&\frac{2RTQ}{\sqrt{a}}\left(12a\dot{a}f_{_{R}}
+6a^2(\dot{R}f_{_{RR}}+\dot{T}f_{_{RT}}+\dot{Q}f_{_{RQ}})
-6a\dot{a}f_{_{Q}}(2\rho_m+a\rho_m')\right.\\\nonumber&-&\left.3a^2\rho_m
(\dot{R}f_{_{RQ}}+\dot{T}f_{_{TQ}}+\dot{Q}f_{_{QQ}})\right)
+\frac{12a^2RT^3Q^2}{a^{\frac{11}{6}}(c_1a^{\frac{1}{3}}+a
T)}(6a^2\dot{a}f_{_{RT}}-3a^2\dot{a}\rho_mf_{_{TQ}})
\\\nonumber&-&\frac{6a^2RT^2Q^2}{a^{\frac{5}{2}}(c_1a^{\frac{1}{3}}+a
T)}(6a^2\dot{a}f_{_{RQ}}-3a^2\dot{a}\rho_mf_{_{QQ}}),\\\nonumber
Y_3&=&\frac{RT}{a^{\frac{7}{6}}}\partial_{_{a}}
+\frac{3RT^2Q}{c_1a^{\frac{1}{3}}+a T}
\left(\frac{2T}{\sqrt{a}}\partial_{_{T}}
-\frac{1}{a^{\frac{1}{3}}}\partial_{_{Q}}\right),\\\nonumber
I_3&=&\frac{RT}{a^{\frac{7}{6}}}\left(12a\dot{a}f_{_{R}}
+6a^2(\dot{R}f_{_{RR}}+\dot{T}f_{_{RT}}+\dot{Q}f_{_{RQ}})
-6a\dot{a}f_{_{Q}}(2\rho_m+a\rho_m')\right.\\\nonumber&-&\left.3a^2\rho_m
(\dot{R}f_{_{RQ}}+\dot{T}f_{_{TQ}}+\dot{Q}f_{_{QQ}})\right)
+\frac{6RT^3Q}{a^{\frac{1}{2}}(c_1a^{\frac{1}{3}}+a
T)}(6a^2\dot{a}f_{_{RT}}-3a^2\dot{a}\rho_mf_{_{TQ}})
\\\nonumber&-&\frac{3RT^2Q}{a(c_1a^{\frac{1}{3}}+a
T)}(6a^2\dot{a}f_{_{RQ}}-3a^2\dot{a}\rho_mf_{_{QQ}}),\\\nonumber
Y_4&=&\left(\frac{7+9a^{\frac{2}{3}}RT}{c_1a^{\frac{1}{3}}+a
T}\right)\left[\frac{RT^2}{3a^{\frac{11}{6}}}\partial_{_{T}}
-\frac{RT}{6a^{\frac{5}{2}}}\partial_{_{Q}}\right],\quad
I_4=\frac{RT^2(7+9a^{\frac{2}{3}}RT)}{3a^{\frac{11}{6}}
(c_1a^{\frac{1}{3}}+aT)}\\\nonumber&\times&(6a^2\dot{a}f_{_{RT}}-3a^2\dot{a}\rho_mf_{_{TQ}})
-\frac{RT(7+9a^{\frac{2}{3}}RT)}{6a^{\frac{5}{2}}
(c_1a^{\frac{1}{3}}+aT)}(6a^2\dot{a}f_{_{RQ}}-3a^2\dot{a}\rho_mf_{_{QQ}}),\\\nonumber
Y_5&=&\left(\frac{14a^{\frac{2}{3}}Q+18a^{\frac{4}{3}}RT}{c_1a^{\frac{1}{3}}+a
T}\right)\left[\frac{RT^2}{3a^{\frac{11}{6}}}\partial_{_{T}}
-\frac{RT}{6a^{\frac{5}{2}}}\partial_{_{Q}}\right],\quad
I_5=\frac{2RT^2Qa^{\frac{2}{3}}(7+9a^{2}RT)}{3a^{\frac{11}{6}}
(c_1a^{\frac{1}{3}}+aT)}\\\nonumber&\times&(6a^2\dot{a}f_{_{RT}}-3a^2\dot{a}\rho_mf_{_{TQ}})
-\frac{a^{\frac{2}{3}}RTQ(7+9a^{\frac{2}{3}}RT)}{3a^{\frac{5}{2}}
(c_1a^{\frac{1}{3}}+aT)}(6a^2\dot{a}f_{_{RQ}}-3a^2\dot{a}\rho_mf_{_{QQ}}).
\end{eqnarray}
The symmetry generator $Y_1$ identifies scaling symmetry whereas
associated conserved integral $I_1$ defines conservation of linear
momentum. In the present case, it is too difficult to evaluate
explicit form of the scale factor for the established exponential
$f(R,T,Q)$ model.

\section{Final Remarks}

The non-minimally coupled gravitational theories put forward
fascinating approaches to explore evolutionary phases of the cosmos
from its origin to the current state. This work determines exact
solution of flat and isotropic cosmos filled with dust distribution
via Noether symmetry scheme in $f(R,T,Q)$ theory. Following this
approach, a system with non-zero boundary term ($\mathcal{B}\neq0$)
preserves Noether gauge symmetry while dynamical system with zero
boundary term ($\mathcal{B}=0$) recovers simple Noether point
symmetry. These symmetries are categorized into translational and
spatial symmetries identifying corresponding conservation laws,
i.e., energy and linear/angular momentum conservation, respectively.

In order to discuss the impact of weak or strong interactions
between curvature invariant and matter variables on
geodesic/non-geodesic test particles, we have considered different
non-minimally coupled models such as $f(R,Q)$, $f(R,T)$ and $f(T,Q)$
models. For each choice and general $f(R,T,Q)$ model, we have found
scaling symmetry along conservation of linear momentum for
$f(R,Q),~f(T,Q)$ and $f(R,T,Q)$ models. For $f(R,T)$ model, the
system of over-determining equations fail to produce scaling
symmetry and relative conservation law. For each considered model,
the explicit forms of energy density and generic function are
calculated that yield exact solutions for scale factor whereas for
general $f(R,T,Q)$ model, the formulation of exact solution is not
possible. We have studied the cosmological nature of these solutions
via graphical analysis of some standard parameters, i.e., Hubble,
deceleration and effective EoS parameters. According to Planck 2018
constraints, the suggested values of $H$ at $68\%CL$ are given by
\cite{b8}
\begin{eqnarray*}\nonumber
H_0&=&67.27\pm0.60\quad\text{(TT+TE+EE+low E)},\\\nonumber
H_0&=&67.36\pm0.54\quad\text{(TT+TE+EE+low E+lensing)},\\\nonumber
H_0&=&67.66\pm0.42\quad\text{(TT+TE+EE+low E+lensing+BAO)}.
\end{eqnarray*}

Furthermore, we have investigated the existence of physically viable
and stable state of new $f(R,Q),~f(R,T),~f(T,Q)$ and $f(R,T,Q)$
models through the squared speed of sound and well-known
Dolgov-Kawasaki viability conditions. The matter distribution is
also analyzed using fractional densities relative to pressureless
fluid and DE. The observational constraints on $\Omega_m$ and
$\Omega_d$ with $68\%$CL are given as \cite{b8}
\begin{eqnarray}\nonumber
&&\Omega_m=0.3166\pm0.0084,\quad\Omega_d=0.7116\pm0.0084,\\\nonumber&&\text{(TT+TE+EE+low
E)},\\\nonumber
&&\Omega_m=0.3158\pm0.0073,\quad\Omega_d=0.6847\pm0.0073,\\\nonumber&&\text{(TT+TE+EE+low
E+lensing)},\\\nonumber
&&\Omega_m=0.3111\pm0.0056,\quad\Omega_d=0.68889\pm0.0056,\\\nonumber&&\text{(TT+TE+EE+low
E+lensing+BAO)}.
\end{eqnarray}
The consistency of above mentioned parameters is checked against
recent observational data of Planck 2018. The results are summarized
as follows.
\begin{itemize}
\item \textbf{$f(R,Q)$ Model}
\end{itemize}
In the absence of boundary term, the cosmological parameters
identify a transition from accelerated to decelerated expansion
while the stable as well as viable $f(R,Q)$ model describes phantom
and quintessence regions. The fractional densities meet with Planck
limitations whereas the consistency disturbs as time passes due to
the dominance of matter fractional density. Zubair et al. \cite{r1}
investigated cosmic evolution using particular $f(R,Q)$ model and
power-law Hubble parameter without using Noether Symmetry approach.
They constructed model constraints that refer to $\Lambda$CDM limit
and explain current accelerated expansion.
\begin{itemize}
\item \textbf{$f(R,T)$ Model}
\end{itemize}
For this choice of model, the cosmological analysis characterizes a
smooth exit from quintessence to phantom DE era. The stable and
viable non-minimally coupled $f(R,T)$ model is found to be
consistent with Chaplygin gas model. The graphical evaluation of
fractional densities supports accelerated expansion of the universe.
Momeni et al. \cite{a3} used simple Noether symmetry technique to
evaluate symmetry generators and conserved integral in $f(R,T)$
theory. They considered minimally interacting $f(R,T)$ model and
determined exact solution corresponding to decelerated expansion of
the cosmos whereas we have found non-minimal $f(R,T)$ model
specifying accelerated cosmic expansion. In this regard, it is
interesting to mention that current accelerated expansion can
successfully be discussed in the presence of non-minimal
interactions of scalar curvature and matter contents.
\begin{itemize}
\item \textbf{$f(T,Q)$ Model}
\end{itemize}
The graphical study of cosmological parameters leads to accelerated
expansion of the universe for evaluated Noether point symmetries.
The minimally coupled viable $f(T,Q)$ model is found to be
consistent with Chaplygin gas model whereas the model becomes
unstable initially but recovers a stable state as time grows. The
fractional densities are consistent with accelerated expanding
cosmos for zero boundary term.
\begin{itemize}
\item \textbf{$f(R,T,Q)$ Model}
\end{itemize}
For the generalized non-minimally coupled model, we have found an
explicit form of energy density and exponential $f(R,T,Q)$ model. In
this case, we have determined five symmetry generators out of which
only one identifies scaling symmetry with linear conservation of
momentum. Due to strong non-minimal interactions between
curvature-matter variables, it is not possible to measure
cosmological solution.

Sharif and Gul \cite{cm1} examined some physically viable
anisotropic solutions through the Noether symmetry technique in
$f(R, T^2)$ theory. They considered a minimally coupled $f(R, T^2)$
model to determine exact solutions and examined the behavior of
solutions via some cosmological parameters. For Bianchi type III
model, the graphical behavior of cosmological parameters represent
accelerated expansion whereas analysis of fractional densities does
not preserve this consistency for the same model. In case of
Kantowski-Sachs model, the parameters identify decelerated expansion
while DE fractional density dominates matter fractional density. In
the present study, we have formulated non-minimally coupled
isotropic exact solutions. For non-minimal geodesic $f(R,Q)$ model,
the cosmological parameters identify a transition from accelerated
to decelerated expansion whereas the fractional densities analysis
shows dominance of DE initially and this dominance disturbs as time
passes due to increasing amount of matter fractional density. For
constructed non-geodesic $f(R,T)$ model, the parameters characterize
a smooth exit from quintessence to phantom DE era while fractional
densities support accelerated expansion of the universe. In case of
minimal non-geodesic $f(T,Q)$ solution, the study of cosmological
parameters and fractional densities lead to accelerated expansion of
the universe. Thus, it is worth notifying that the analysis of
present cosmological solutions and fractional densities are
compatible with each other.

Finally, We have found scaling symmetry generator and linear
momentum conservation for each case except for $f(R,T)$ model, where
symmetries and respective conservation laws do not correspond to any
standard symmetry or conserved entity. The new non-minimally coupled
models are found to be stable and viable in the background of
pressureless fluid. These models also preserve compatibility with
Chaplygin gas model, quintessence and phantom regions. It is
interesting to conclude that the isotropic cosmological solutions
interpret accelerated cosmic expansion whenever generic function
involves non-minimal coupling. It would be interesting to explore
the impact of non-minimal curvature-matter coupling in the
background of anisotropic universe models. Using Noether symmetry
approach, the study of cosmological configurations like black hole
and wormhole would be much fascinating.\\\\
\textbf{Data Availability Statement:} No new data were created or
analyzed in this study.

\end{document}